\documentclass [letter, 12pt]{article}
\usepackage{a4wide, amsmath, amsfonts, amssymb, geometry}

\usepackage{bm}
\usepackage{fancyhdr, dsfont}
\usepackage[latin1]{inputenc}
\usepackage{graphicx}
\usepackage{float}
   
\restylefloat{figure}
\usepackage{epstopdf}

\newcommand{\nwc}{\newcommand}
\nwc{\ba}  {\begin{array}}
\nwc{\ea}  {\end{array}}
\nwc{\bdm} {\begin{displaymath}}
\nwc{\edm} {\end{displaymath}}

\nwc{\bea} {\begin{equation}\ba{lcl}}
\nwc{\eea} {\ea\end{equation}}

\nwc{\bda} {\bdm\ba{lcl}} 
\nwc{\eda} {\ea\edm}

\nwc{\bc}  {\begin{center}}
\nwc{\ec}  {\end{center}}

\nwc{\ds}  {\displaystyle}
\nwc{\bmat}{\left(\ba}
\nwc{\emat}{\ea\right)}
\nwc{\nn}  {\nonumber}
\nwc{\nnn} {\nonumber \vspace{.2cm} \\ }
\nwc{\ra}  {\rightarrow}
\nwc{\lra} {\longrightarrow}

\nwc{\p} {\partial}

\def\beq{\begin{equation}}
\def\eeq{\end{equation}}

\newcommand{\vecb}{\left(\begin{array}{c}}
\newcommand{\vece}{\end{array}\right)}
\newcommand{\ccb}{\left(\begin{array}{cc}}
\newcommand{\cce}{\end{array}\right)}
\newcommand{\cccb}{\left(\begin{array}{ccc}}
\newcommand{\ccce}{\end{array}\right)}
\newcommand{\ccccb}{\left(\begin{array}{cccc}}
\newcommand{\cccce}{\end{array}\right)}
\newcommand{\cccccb}{\left(\begin{array}{ccccc}}
\newcommand{\ccccce}{\end{array}\right)}

\newcommand{\pa}{\partial}

\newcommand{\al}{\alpha}
\newcommand{\be}{\beta}
\newcommand{\ga}{\gamma}
\newcommand{\de}{\delta}

\newcommand{\vep}{\varepsilon}

\newcommand{\si}{\sigma}
\newcommand{\la}{\lambda}
\newcommand{\ka}{\kappa}

\newcommand{\bi}{\bar\imath}

\newcommand{\Ga}{\Gamma}

\newcommand{\te}{\textrm}
\newcommand{\eq}{ \ \ = \ \ }
\newcommand{\co}{\ , \ \ \ \ \ \ }
\newcommand{\dd}{\mathrm{d}}
\newcommand{\ee}{\mathrm{e}}

\newcommand{\rhob}{\bar \rho}

\newcommand{\dal}{\dot{\alpha}}
\newcommand{\dbe}{\dot{\beta}}
\newcommand{\dga}{\dot{\gamma}}
\newcommand{\dde}{\dot{\delta}}

\newcommand{\RR}{\mathbb R}

\newcommand{\NN}{\mathbb N}

\newcommand{\gab}{\bar{\gamma}}

\newcommand{\ap}{\alpha'}

\allowdisplaybreaks

\begin{document}

\title{\textbf{Higher Spin Scattering in Superstring Theory}\\[0.5cm]}
\author{O. Schlotterer}
\date{\today}
\smallskip
\maketitle
\centerline{\it Max--Planck--Institut f\"ur Physik}
\centerline{\it Werner--Heisenberg--Institut}
\centerline{\it 80805 M\"unchen, Germany}

\medskip\bigskip\vskip2cm \abstract{\noindent We compute scattering amplitudes of leading Regge trajectory states in open superstring theories. Highest spin states at mass level $n$ with spin $s=n+1$ for bosons and $s=n+\frac{1}{2}$ for fermions are generated by particularly simple vertex operators. Hence, the cubic couplings of bosons and fermions on the leading Regge trajectory are given for arbitrary $n$. The same can be achieved for higher point amplitudes, and this article focuses on four point level with one heavy maximum spin state and three massless states in any bose-fermi combination, putting particular emphasis on manifest cyclic symmetry. Except for the four fermion coupling, all our results remain valid in any $D<10$ dimensional compactification scenario, so they might become relevant at LHC in case of an experimentally accessible low string scale. But even if not directly observable, superstring amplitudes provide important clues on higher spin dynamics and their consistent interactions in field theory.}

\vskip4cm
\begin{flushright}
{\small MPP-2010-142}
\end{flushright}

\thispagestyle{empty}

\newpage
\setcounter{tocdepth}{2}
\tableofcontents

\numberwithin{equation}{section}

\newpage

\section{Introduction}
\label{sec:Introduction}

One of the most remarkable features of string theory is the presence of an infinite tower of massive vibration modes in its spectrum due to the extended nature of strings. These heavy states are referred to as Regge excitations and their maximum spins are linearly related to their mass squares via $s_{\te{max}} = 1 + \ap m^2$, with $\ap$ denoting the inverse string tension. In other words, string theory requires a cornucopia of higher spin states for its consistency. The research areas of string theory and higher spin field theory strongly support and motivate each other:

\medskip
First of all, the high energy regime of string theory (i.e. the low tension limit $\ap \rightarrow \infty$ where Regge excitations formally become massless) provides a fruitful laboratory to learn about higher spin gauge theory. Typical no-go theorems on massless higher spin theories are based on subtle assumptions that may well prove too restrictive, such as finiteness of the spectrum or minimal couplings. Results from string theory naturally have the required properties to bypass these no-go theorems.

\medskip
Secondly, taking all the higher spin modes into account might pave the way beyond the on-shell first quantized picture of string theory and shed light on its true quantum degrees of freedom. Moreover, ideas have been around that masses of Regge excitations arise from some sort of generalized Higgs effect for spontaneous breaking of higher spin gauge symmetries. From this viewpoint, string theory itself might well prove to be a major motivation to study higher spin theories.

\medskip
Therefore, investigating higher-spin dynamics will help to better understand string theory and, vice versa, a closer look at string theory at high energies in the $\ap \rightarrow \infty$ limit can provide important clues on higher-spin dynamics. After first explicit investigation of higher spin interactions in \cite{B1,B2,B3,B4,B5}, great progress in finding all order vertices for totally symmetric tensors was made in \cite{Vasiliev:1990en,Vasiliev:2003ev}, see \cite{Bekaert:2005vh} for a review beyond cubic level.


\medskip
The main motivation for this article is to extend the analysis of \cite{bos,Taronna:2010qq} on higher spin interactions in bosonic string theory to the open superstring. The authors investigated cubic and quartic couplings of highest spin $s= n+1$ states at mass level $n$ with $m^2 = \frac{n}{\ap}$ on the bosonic string. They extracted their massless limits, explained implications on higher spin gauge symmetry and identified the underlying currents which become conserved in the $\ap \rightarrow \infty$ limit. Recently, similar calculations have been carried out in heterotic string theories \cite{Bianchi:2010es} where stable higher spin BPS states were shown to exist and a large class of three- and four point scattering amplitudes was computed for states with higher spin content on the bosonic side of the heterotic string. Further interesting results on scattering of exotic massless higher spin states from a decoupled sector of superstring theory can be found in recent work \cite{Polyakov:2009pk,Polyakov:2010sk}.

\medskip
In addition to the on-shell amplitudes from the aforementioned references, off shell vertices in bosonic string theory are constructed in the updated version of \cite{bos} and in \cite{Fotopoulos:2010ay}. Generalizations of the generating functionals involved, which directly lead to off-shell amplitudes, can be found in \cite{Manvelyan:2010je}. The unique off-shell cubic vertices generated by this generating function were derived earlier in \cite{Manvelyan:2010jr} from the point of view of higher spin gauge field theory.


\medskip
In this article, we will scatter massive higher spin states of open superstring theories, limiting our attention to the leading Regge trajectory of the superstring spectrum with spin $s= n+1$ for bosons and $s=n+\frac{1}{2}$ for fermions at mass level $n$. It is technically very difficult to go beyond leading Regge trajectory: Although the full superstring spectrum was recently organized into explicitly known $SO(9)$ content \cite{Hanany:2010da}, no superstring vertex operators with general mass level $n$ have been fully worked out for subleading Regge trajectories. 

\medskip
Knowledge of highest spin three- and four-point interactions can prove valuable for factorization properties of massless multileg superstring disk amplitudes. The complexity of formfactors in multi gluon scattering grows enormously with increasing number of external legs, see \cite{OP, Stieberger:2006te, Stieberger:2007jv, Stieberger:2007am} for six- and seven-point results. But some promising expansion methods bring them into a form which encourages to disentangle the contributions of individual spins. The simplifications on the basis of this spin separation might in the end pave the way for new recursion relations.

\medskip
Scattering amplitudes of higher spin states may even be relevant for phenomenology. As pointed out in \cite{D1,D2}, the string scale can be as low as a few TeV provided that some of the extra dimensions are sufficiently large. Experimental signatures of Regge excitations at accelerators such as the LHC were firstly discussed for low string scale scenarios in \cite{PP,lhc1,lhc2}. In these references, resonances in lepton, quark- and gluon scattering processes due to exchange of virtual higher spin states were predicted. Once the mass threshold $M_{\te{string}} = \ap^{-1/2}$ is crossed in the center-of-mass energies of colliding partons, one would also see the string resonance states produced directly, in association with jets, photons and other particles \cite{anch}. In subsequent work \cite{lhc3}, production rates of first mass level excitations of quarks and gluons were computed. The amplitudes presented in this paper can also be regarded as a generalization of this mass level one discussion. In fact, gluon fusion and quark antiquark annihilation can produce any leading trajectory boson if the mass threshold is reached in the center of mass energies of the colliding partons, and similarly, higher spin fermions appear in quark gluon scattering at sufficiently high energy.

\medskip
This article is organized as follows: In the following section \ref{sec:vertex}, the vertex operators for leading trajectory states are reviewed, some notation is fixed and the applicability to compactifications is explained. Then, three point couplings of highest spin states at generic mass levels $n_1,n_2$ and $n_3$ are computed and discussed in section \ref{sec:3}. Finally, section \ref{sec:4pt1} is dedicated to four point amplitudes of higher spin bosons and fermions to their massless cousins with potential relevance for dijet parton scattering processes at LHC.

\section{Vertex operators of leading Regge trajectory states}
\label{sec:vertex} 

\subsection{Higher spin bosons (NS sector)}

States of maximum spin $s=n+1$ at mass level $n$ are obtained by applying one worldsheet fermion oscillator $\psi^\nu_{-1/2}$ and $n$ boson oscillators $\al_{-1}^{\mu_i}$ to the NS ground state $| 0 \rangle_{\te{NS}}$, the spacetime indices being contracted with a totally symmetric tensor $\phi_{\mu_1 ... \mu_n \nu} = \phi_{(\mu_1 ... \mu_n \nu)}$. In contrast to bosonic string theory, vertex operators of the superstring carry a superghost charge to take gauge fixing of superconformal invariance on the worldsheet into account. The NS highest spin state $\psi_{-1/2} (\al_{-1})^n | 0 \rangle$ is generated by the following vertex operator in its ``canonical'' ghost picture of ghost charge $-1$:
\beq
V_n^{(-1)}(\phi,k) \eq \frac{1}{\sqrt{2\ap}^n} \; (T^a) \, \phi_{\mu_1 ... \mu_n \nu} \, : \, i\pa X^{\mu_1} \, ... \, i \pa X^{\mu_n} \, \psi^\nu \, e^{-\phi} \, e^{ikX} \, :
\label{2,1}
\eeq
The Chan Paton matrix $T$ is a generator of the gauge group, its adjoint index $a$ represents the colour degrees of freedom of the state created by (\ref{2,1}).

\medskip
BRST invariance requires the totally symmetric rank $s$ wave function $\phi$ to satisfy the massive Pauli-Fierz conditions
\beq
k^{\mu_i} \, \phi_{\mu_1 ... \mu_s} \eq \eta^{\mu_i \mu_j} \, \phi_{\mu_1 ... \mu_s} \eq 0 \ .
\label{2,2}
\eeq
The requirement of transversality and vanishing trace projects $\phi$ into the irreducible spin $n+1$ representation of the little group $SO(1,D-2)$.

\medskip
In order to cancel the background ghost charge on a genus $g$ Riemann surface, a $g$--loop superstring amplitude needs an overall superghost charge of $2g-2$, i.e. total ghost charge $-2$ at tree level. Therefore, already a three point disk amplitude requires a higher ghost picture analogue of the vertex (\ref{2,1}). By application of the picture raising operator $P_{+1} = \frac{1}{\sqrt{2\ap}} \, i\pa X_\mu \psi^\mu e^{\phi}$, we obtain the following zero picture vertex for the highest spin state of mass level $n$:
\begin{align}
V_n^{(0)}(\phi,k) \eq \frac{1}{\sqrt{2\ap}^{n-1}} \; (T^a) \, &\phi_{\mu_1 ... \mu_n \nu} \, : \, \left( \, n \, \pa \psi^{\mu_n} \, \psi^\nu \ + \ (k \cdot \psi) \, \psi^\nu \, i \pa X^{\mu_n} \ + \ \frac{1}{2\ap} \; i \pa X^{\mu_n} \, i \pa X^{\nu} \, \right) \notag \\
& \times \, i \pa X^{\mu_1} \, ... \, i \pa X^{\mu_{n-1}} \, e^{ikX} \, :
\label{2,3}
\end{align}
Note that the first terms is absent for the massless gauge boson at $n=0$.

\subsection{Higher spin fermions (R sector)}

The spin $s=n+\frac{1}{2}$ state on the leading fermionic trajectory at mass level $n$ belongs to the R sector of the superstring and therefore involves spin fields $S_\al$ which generate the $SO(1,9)$ Weyl spinor of R ground states. Apart from the usual $(\al_{-1})^n$ oscillators coupled to a left handed spinor wave function $v$, fermions at mass level $n$ additionally require a $\psi_{-1} \psi_0 (\al_{-1})^{n-1}$ contribution along with a right handed tensor spinor $\rhob$ to ensure BRST invariance \cite{vanP,ferm}. The vertex operator in its canonical ghost picture carries charge $-\frac{1}{2}$:
\begin{align}
V_{n}^{\left(-\frac{1}{2} \right)}(v, \rhob ,k) \eq & : \, \bigl[ \, v_{\mu_1 ... \mu_{n}}^{\al} \, i\pa X^{\mu_1} \, ... \, i \pa X^{\mu_n} \ + \ 2\al' \, \rhob^{\mu_1 ... \mu_{n-1} \nu}_{\dbe} \, i \pa X_{\mu_1} \, ... \, i \pa X_{\mu_{n-1}} \, \psi_{\nu} \, \psi^\la \, \gab_\la^{\dbe \al } \, \bigr] \notag \\
& \ \ \ \ \ \ \ \ \ \ \ \ \ \times \ \frac{1}{\sqrt{2\ap}^n} \; (T^a) \, S_{\al} \, e^{-\phi/2} \, e^{ikX} \, :
\label{3,1}
\end{align}
The irreducible spin $n+\frac{1}{2}$ component of the wave function $v$ is traceless with respect to both $\eta^{\mu_i \mu_j}$ and  $\ga^{\mu_i}_{\al \dbe}$. Together with the on-shell constraint, we have 
\beq
k^{\mu_i} \, v_{\mu_1 ... \mu_{n}}^{\al} \eq \eta^{\mu_i \mu_j} \, v_{\mu_1 ... \mu_{n}}^{\al} \eq   v_{\mu_1 ... \mu_{n}}^{\al} \,  \ga_{\al \dbe}^{\mu_i} \eq 0
\label{3,2}
\eeq
for the left handed part of the polarization, and the right handed counterpart $\rhob$ is determined by a massive Dirac equation (with Feynman slash notation $\ga^\nu_{\al \dbe} k_\nu \equiv \not \! k_{\al \dbe}$ and $\gab_\nu^{\dal \be} k^{\nu} \equiv \not \! k^{\dal \be}$)
\beq
v_{\mu_1 ... \mu_n}^{\al}  \, \ga^\nu_{\al \dbe}\, k_\nu \ \ \equiv \ \ v_{\mu_1 ... \mu_n}^{\al}  \, \not \! k_{\al \dbe} \eq - \, (D-2) \, (\rhob_{\mu_1 ... \mu_n})_{\dbe} \ ,
\label{high7a}
\eeq
the $\rhob$ wavefunction of course inherits the properties (\ref{3,2}) of $v$.

\medskip
Although we are mostly interested in the $D=10$ dimensional result, we will keep the number $D$ of dimensions general in describing massive fermions, see the following subsection as well as appendix \ref{sec:D}.

\medskip
For the purpose of a more streamlined notation, the $h = 1 + \frac{D}{16}$ combination of $\psi \psi S$ fields in the $\rhob$ part of the fermion vertex deserves its own shorthand
\beq
K^{\dal}_\mu \ \ := \ \ \psi_\mu \, \psi^\nu \, \gab^{\dal \be}_\nu \, S_\be \ \ \equiv \ \ \psi_{\mu} \not \! \psi^{\dal \be} \, S_{\be}  \ .
\label{high9}
\eeq
In appendix \ref{appB}, we will give the relevant correlation functions for $K_\mu^{\dal}$ rather than for $\psi^\mu \psi^\nu S_{\al}$. The spin $n+\frac{1}{2}$ vertex operator then takes the following form:
\begin{align}
V_{n}^{\left(-\frac{1}{2} \right)}(v, \rhob ,k) \eq & : \, \bigl[ \, v_{\mu_1 ... \mu_{n}}^{\al} \, i\pa X^{\mu_1} \, ... \, i \pa X^{\mu_n} \, S_{\al} \ + \ 2\al' \, \rhob^{\mu_1 ... \mu_{n-1} \nu}_{\dbe} \, i \pa X_{\mu_1} \, ... \, i \pa X_{\mu_{n-1}} \, K_\nu^{\dbe} \, \bigr] \notag \\
& \ \ \ \ \ \ \ \ \ \ \ \ \ \times \ \frac{1}{\sqrt{2\ap}^n} \; (T^a) \, e^{-\phi/2} \, e^{ikX} \, :
\label{high10}
\end{align}
The fact that $K_\nu^{\dbe}$ is contracted by a $\ga$ traceless wavefunction $\bar \rho$ will simplify the CFT correlators, see appendix \ref{appB}.

\subsection{Applicability in lower dimensions}
\label{sec:dim}

Although we have adapted our whole setup to full-fledged ten dimensional superstring theory with spacetime filling D9 branes, most of the results presented in this paper can be taken over to lower dimensional D$p$ brane worldvolumes and compactification geometries. 

\medskip
Dimensional reduction of bosonic states and their couplings is straightforward: Suppose $\phi_{\mu_1...\mu_s}$ with $\mu_i = 0,...,9$ is the initially $SO(1,9)$ covariant wave function of the $s=n+1$ state on the leading trajectory, then compactification of $10-D$ spacetime dimensions breaks Lorentz symmetry to $SO(1,D-1)$ and hence requires to specify for each index $\mu_i$ whether it belongs to the $D$ uncompactified dimensions or to the internal ones. Let us denote the $SO(1,D-1)$ indices by $m_i$, then only the $\phi_{m_1...m_s}$ components of the ten dimensional polarization tensor describes a maximum spin state from the $D$ dimensional point of view. Neglecting Kaluza Klein excitations, we can reduce the ten dimensional momenta to $k_\mu = (k_m,0)$ with a $10-D$ component zero vector in the compactified directions. The only modification of the $s=n+1$ vertex operator (\ref{2,1}) upon compactifying $10-D$ dimensions is the replacement of $SO(1,9)$ indices $\mu_i$ by $SO(1,D-1)$ indices $m_i$. Correlation functions involving exclusively the spacetime fields $\psi^\mu$ and $i\pa X^\mu$ from the highest spin vertex (\ref{2,1}) do not depend on the range of their indices, see appendix \ref{appA}, so the result (\ref{2,7}) and in fact any higher-point bosonic highest spin state amplitude cannot depend on $D$.

\medskip
Dimensional reduction of spinorial fields is way more subtle and also model dependent, that is why the detailed discussion is moved to appendix \ref{sec:D}. The important message is the following: All the two fermion amplitudes presented in this article, namely
\beq
\langle V_{n_1}^{(-1)}(\phi) \, V_{n_2}^{(-\frac{1}{2})}(v_2) \, V_{n_3}^{(-\frac{1}{2})}(v_3) \rangle \co \begin{array}{c}
\langle V_{0}^{(0)}(\xi_1) \, V_{0}^{(-1)}(\xi_2) \, V_{0}^{(-\frac{1}{2})}(u) \, V_{n}^{(-\frac{1}{2})}(v) \rangle \\
\langle V_{0}^{(-\frac{1}{2})}(u_1) \, V_{0}^{(0)}(\xi) \, V_{0}^{(-\frac{1}{2})}(u_3) \, V_{n}^{(-1)}(\phi) \rangle \end{array} \ ,
\label{2fermi}
\eeq
turn out to be independent on $D$ (apart from the two fermions' relative chirality) -- although this is not obvious during the computation before eliminating all the $\rhob$ via (\ref{high7a}). The basic reason for this lies in the decoupling of the internal CFT from the bosonic highest spin vertex. Hence, two fermion amplitudes involve the conformal fields carrying the compactification details in a two point function only, which is fixed up to normalization in any CFT. This argument obviously fails at the level of four fermions. For details, we refer the reader to appendix \ref{sec:D}.

\medskip
For $D=4$, this property is in lines with the observation of universality in \cite{lhc1,lhc2,lhc3}. In these references, color stripped two fermion amplitudes were shown to be the same for adjoint fermions from the untwisted CFT sector and for chiral fermions located at brane intersections, realized by boundary changing conformal fields.

\section{Higher spin three point couplings}
\label{sec:3}

Having introduced the highest spin vertex operators in various ghost pictures, we are now ready to compute three point amplitudes for arbitrary mass levels $n_{i}$ and spins $s_{i}= n_i + 1$ of $s_i = n_i + \frac{1}{2}$, for bosons and fermions respectively. More precisely, we will compute their color stripped subamplitudes ${\cal A}_{(s_1,s_2,s_3)}$ associated with the Chan Paton trace $\te{Tr} \bigl\{ T^{a_1} \, T^{a_2}\, T^{a_3} \bigr\}$ in the canonical ordering $(1,2,3)$ of the external legs. The $s_i$ subscripts of the amplitude indicate the spins involved.

\medskip
In order to put the two (cyclically) inequivalent orderings $(1,2,3)$ and $(1,3,2)$ together, we have to weight the associated traces with the overall worldsheet parity $(-1)^{1+N}$ of the three states with overall mass level $N=n_1+n_2+n_3$. This gives rise to either the structure constants $f^{a_1 a_2 a_3}$ of the gauge group (defined by commutators $\bigl[ T^a , T^b \bigr] = i f^{abc} T_c$) or the symmetrized three trace $d^{a_1 a_2 a_3} := \te{Tr} \bigl\{ T^{(a_1} \, T^{a_2}\, T^{a_3)} \bigr\}$ as the resulting color factor:
\beq
{\cal A}^{\te{full}}_{(s_1 ,s_2 ,s_3)} \eq \left\{ \, \begin{array}{rl} \frac{i}{2} \; f^{a_1 a_2 a_3} \, {\cal A}_{(s_1 ,s_2 ,s_3)} &: \ N \ \te{even} \\ 2 \, d^{a_1 a_2 a_3} \, {\cal A}_{(s_1 ,s_2 ,s_3)} &: \ N \ \te{odd} \end{array} \right.
\label{color}
\eeq
The generators are normalized such that $\te{Tr} \bigl\{ T^{a} \, T^{b} \bigr\} = \de^{ab} / 2$.

%

\subsection{Three boson couplings}
\label{sec:3pt}

The general prescription for disk amplitudes gives us the following expression for the scattering of three bosons on the leading Regge trajectory\footnote{In slight abuse of notation, we neglect that the vertex operators $V_{n_i}$ contain Chan Paton generators $T^{a_i}$.}:
\begin{align}
{\cal A}&_{(s_1,s_2,s_3)} \ \ = \ \ \langle \, cV_{n_1}^{(0)}(\phi^1,k_1) \, cV_{n_2}^{(-1)}(\phi^2,k_2) \, cV_{n_3}^{(-1)}(\phi^3,k_3) \, \rangle \notag \\
&= \ \ \sqrt{2\ap}^{1-n_1-n_2-n_3} \, \langle \, c(z_1) \, c(z_2) \, c(z_3) \, \rangle \, \langle \, e^{-\phi(z_2)} \, e^{-\phi(z_3)} \, \rangle \, \phi^1_{\mu_1 ... \mu_{n_1} \rho} \, \phi^2_{\nu_1 ... \nu_{n_2} \tau} \, \phi^3_{\la_1 ... \la_{n_3} \si} \notag \\
& \ \ \ \ \times \ \biggl\{ \, \langle \, i\pa X^{\mu_1} \, ... \, i \pa X^{\mu_{n_1-1}}(z_1) \, i\pa X^{\nu_1} \, ... \, i \pa X^{\nu_{n_2}}(z_2) \, i\pa X^{\la_1} \, ... \, i \pa X^{\la_{n_3}}(z_3) \, \prod_{j=1}^3 e^{ik_j X(z_j)} \, \rangle \, \biggr. \notag \\
& \ \ \ \ \ \ \ \ \ \ \biggl. \ \times \ n_1 \, \langle \, \pa \psi^{\mu_{n_1}}(z_1) \, \psi^{\rho}(z_1) \, \psi^{\tau}(z_2) \, \psi^{\si}(z_3) \, \rangle \, \biggr. \notag \\
& \ \ \ \ \ \ \ \ \ \ \biggl. + \ \langle \, i\pa X^{\mu_1} \, ... \, i \pa X^{\mu_{n_1}}(z_1) \, i\pa X^{\nu_1} \, ... \, i \pa X^{\nu_{n_2}}(z_2) \, i\pa X^{\la_1} \, ... \, i \pa X^{\la_{n_3}}(z_3) \, \prod_{j=1}^3 e^{ik_j X(z_j)} \, \rangle \, \biggr. \notag \\
& \ \ \ \ \ \ \ \ \ \ \biggl. \ \times \ k_\al^1 \, \langle \, \psi^{\al}(z_1) \, \psi^{\rho}(z_1) \, \psi^{\tau}(z_2) \, \psi^{\si}(z_3) \, \rangle \biggr. \notag \\
& \ \ \ \ \ \ \ \ \ \ \biggl. + \ \langle \, i\pa X^{\mu_1} \, ... \, i \pa X^{\mu_{n_1}} \, i\pa X^{\rho}(z_1) \, i\pa X^{\nu_1} \, ... \, i \pa X^{\nu_{n_2}}(z_2) \, i\pa X^{\la_1} \, ... \, i \pa X^{\la_{n_3}}(z_3) \, \prod_{j=1}^3 e^{ik_j X(z_j)} \, \rangle \biggr. \notag \\
& \ \ \ \ \ \ \ \ \ \ \biggl. \ \times \ \frac{1}{2 \ap} \; \langle \, \psi^{\tau}(z_2) \, \psi^{\si}(z_3) \, \rangle \, \biggr\} 
\label{2,4}
\end{align}
The ghost- and superghost correlators are given by $\langle c(z_1)  c(z_2)  c(z_3)  \rangle = z_{12} z_{13} z_{23}$ and $\langle e^{-\phi(z_2)}  e^{-\phi(z_3)}  \rangle = z_{23}^{-1}$, the correlation functions involving the NS fermion $\psi$ and the string coordinate $X$ are gathered in the appendix \ref{appA}. By using their explicit form, we find that the $z_{ij}$ dependence drops out in every individual term as required by $SL(2,\RR)$ invariance.

\medskip
Cyclic symmetry of the three point couplings is initially obscured by the asymetric assignment of superghost charges. But the combination of the three terms in (\ref{2,4}) (with some redefinition of summation variables $i,j,k$) makes sure that the end result is independent on the ghost picture choices, i.e. that the labels $1,2,3$ enter on the same footing. After some algebra, one ends up with
\begin{align}
{\cal A}_{(s_1,s_2,s_3)} \ \ &= \ \  \sqrt{2\ap}^{s_1+s_2+s_3} \, n_1 ! \, n_2! \, n_3! \, \sum_{i,j,k \in {\cal I}_b} \, \frac{(2\ap)^{-i-j-k} \, ( i \, s_3  \, + \, j \, s_2  \, + \, k \, s_1 \, - \, ij \, - \, ik \, - \, jk)}{i! \, j! \, k! \, (s_1 - i - j)! \, (s_2 - i - k)! \, (s_3 - j-k)!} \notag \\
& \ \ \ \ \times \, \phi^1_{\mu_1 ... \mu_j} \! \, ^{\nu_1 ... \nu_i} \! \,_{\rho_1 ... \rho_{s_1-i-j}} \, k_2^{\rho_1} \, ... \, k_2^{ \rho_{s_1-i-j}} \,
\phi^2_{\nu_1 ... \nu_i} \! \, ^{\la_1 ... \la_k} \! \,_{\tau_1 ... \tau_{s_2-i-k}} \, k_3^{\tau_1} \, ... \, k_3^{ \tau_{s_2-i-k}} \notag \\
& \ \ \ \ \times \, \phi^3_{\la_1 ... \la_k} \! \, ^{\mu_1 ... \mu_j} \! \,_{\si_1 ... \si_{s_3-j-k}} \, k_1^{\si_1 } \, ... \, k_1^{ \si_{s_3-j-k}} \label{2,5}
\end{align}
where $s_i = n_i+1$ and the summation range ${\cal I}_b$ is defined exactly like in \cite{bos}:
\begin{align}
{\cal I}_b \ \ := \ \ \Bigl\{  \, i,j,k \in \NN_0 \, : \ s_1 - i - j \geq 0 \ , \ \ \ s_2 - i-k \geq 0 \ , \ \ \ s_3 - j - k \geq 0 \, \Bigr\}
\label{2,6}
\end{align}
To lighten our notation, it makes sense to introduce some shorthands. A $k$ fold contraction of tensors $\phi^i$ and $\phi^j$ is denoted by $\phi^i \phi^j \de_{ij}^k = \phi^i_{\mu_1 ... \mu_k} \phi^{j \, \mu_1 ... \mu_k}$, and the momenta which are dotted into the remaining indices of the $\phi$'s are written with their power as a collective index, e.g. $\phi^1 \cdot k_2^p := \phi^1_{\rho_1 ... \rho_p} \, k_2^{\rho_1} ... k_2^{\rho_p}$. This convention admits to rewrite the three point function (\ref{2,5}) more compactly as
\begin{align}
{\cal A}_{(s_1,s_2,s_3)} \ \ &= \ \  \sqrt{2\ap}^{s_1+s_2+s_3} \, n_1 ! \, n_2! \, n_3! \, \sum_{i,j,k \in {\cal I}_b} \, \frac{(2\ap)^{-i-j-k} \, ( i \, s_3  \, + \, j \, s_2  \, + \, k \, s_1 \, - \, ij \, - \, ik \, - \, jk)}{i! \, j! \, k! \, (s_1 - i - j)! \, (s_2 - i - k)! \, (s_3 - j-k)!} \notag \\
& \ \ \ \ \times \, ( \phi^1 \cdot k_2^{s_1-i-j}) \, (\phi^2 \cdot k_3^{s_2-i-k}) \, (\phi^3 \cdot k_1^{s_3-j-k} ) \, \de_{12}^i \, \de_{13}^j \, \de_{23}^k \ .
\label{2,7}
\end{align}
The essential difference to the analogous formula of bosonic string theory (equation (4.1) of \cite{bos}) lies in the factor $i  s_3  \, + \, j s_2  \, + \, k  s_1 \, - \, ij \, - \, ik \, - \, jk$ (where the bosonic string has the universal factor $s_1 s_2 s_3$ instead). As a result, the $i=j=k=0$ term with highest $\ap$ power is absent in superstring theory. For three gauge bosons $s_i = 1$, for instance, the summation range ${\cal I}_b$ admits $(i,j,k) = (0,0,0),\, (1,0,0), \, (0,1,0),\, (0,0,1)$, and suppression of the first term ensures the absence of $F^3$ interactions which were present on the bosonic string.

\medskip
One of the most interesting three point vertices is the electromagnetic coupling of spins $s_2$ and $s_3$, i.e. the special case $n_1=0$ where the first state is set to massless spin $s_1=1$ with polarization vector $\phi^1_\mu \equiv \xi_\mu$. For identical spins $s_2 = s_3 \equiv s$ at level $n=s-1$ we find
\begin{align}
{\cal A}_{(1,s,s)} \ \ &= \ \ (2\ap)^{s+\frac{1}{2}} \, (n!)^2 \, \sum_{k=0}^s \frac{ (2\ap)^{-k} \, \de_{23}^k}{k! \, \bigl[ \, (s-k)! \, \bigr]^2 } \, \biggl\{ \, k \, (\xi \cdot k_2) \, (\phi^2 \cdot k_3^{s-k}) \, (\phi^3 \cdot k_1^{s-k} ) \biggr.
\notag \\
& \ \ \  \biggl. \ + \ \frac{s \, (s-k)}{2\ap} \; (\xi^{\mu} \, k_1^\nu \, - \, \xi^\nu \, k_1^\mu) \, (\phi^2_\mu \cdot k_3^{s-k-1}) \, (\phi^3_\nu \cdot k_1^{s-k-1})  \,  \biggr\} \ ,
\label{2,8alt}
\end{align}
and different spins $s_2, s_3$ give rise to
\begin{align}
{\cal A}_{(1,s_2,s_3)} \ \ &= \ \ \sqrt{2\ap}^{1+s_2+s_3} \, n_2! \, n_3! \, \sum_{k=0}^{\te{min}(s_2,s_3)} \frac{ (2\ap)^{-k} \, \de_{23}^k}{k! \, (s_2-k)! \, (s_3 - k)! } \, \biggl\{ \, k \, (\xi \cdot k_2) \, (\phi^2 \cdot k_3^{s_2-k}) \, (\phi^3 \cdot k_1^{s_3-k} )  \biggr.
\notag \\
& \ \biggl. + \ \frac{s_3 \, (s_2-k)}{2\ap} \; \xi^\mu \, (\phi^2_\mu \cdot k_3^{s_2-k-1}) \, (\phi^3 \cdot k_1^{s_3-k})  \ + \ \frac{s_2 \, (s_3-k)}{2\ap} \;  (\phi^2 \cdot k_3^{s_2-k}) \, \xi^\mu \, (\phi^3_\mu \cdot k_1^{s_3-k-1}) \, \biggr\}  \ .
\label{2,9alt}
\end{align}
As a further specialization, the coupling of two massless gauge bosons with polarization vectors $\xi_2, \xi_3$ to massive higher spin states $\phi$ is found to be
\begin{align}
{\cal A}_{(s,1,1)} \eq \, &\sqrt{2\ap}^{n+1} \; \biggl\{ \, \frac{n}{2\al'} \; \xi_2^\mu \, \xi_3^\nu \, (\phi_{\mu \nu} \cdot k_2^{n-1}) \ + \ (\xi_3 \, k_1) \, \xi_2^{\mu} \, (\phi_\mu \cdot k_2^{n}) \biggr. \notag \\
\biggl. \ \ \ \ \ \ \ \ \ \ \ & \ + \ (\xi_2 \, k_3) \, \xi_3^{\mu} \, (\phi_\mu \cdot k_2^n) \ + \ (\xi_2 \, \xi_3) \, (\phi \cdot k_2^{n+1}) \, \biggr\} \ .
\label{high3}
\end{align}
Note that in contrast to heterotic string theories, higher spin fields always decay into lower spin fields. In the heterotic case at least the BPS higher spin fields are protected from that decay \cite{Bianchi:2010es}.

\subsection{Two fermion, one boson coupling}
\label{sec:3ptf}

A second class of nonvanishing three point couplings among leading Regge trajectory states involves two fermions and one boson. This time, all the vertex operators can appear in their canonical ghost picture:
\begin{align}
{\cal A}&_{(s_1,n_2+\frac{1}{2},n_3+\frac{1}{2})} \ \ = \ \ \langle \, cV_{n_1}^{(-1)}(\phi,k_1) \, cV_{n_2}^{(-\frac{1}{2})}(v^2,\rhob^2,k_2) \, cV_{n_3}^{(-\frac{1}{2})}(v^3, \rhob^3,k_3) \, \rangle \notag \\
&= \ \ \sqrt{2\ap}^{-n_1-n_2-n_3} \, \langle \, c(z_1) \, c(z_2) \, c(z_3) \, \rangle \, \langle \, e^{-\phi(z_1)} \, e^{-\phi(z_2)/2} \, e^{-\phi(z_3)/2} \, \rangle \, \phi_{\mu_1 ... \mu_{n_1} \si}  \notag \\
& \ \  \times \, \biggl\{ \, v^{2 \ \al}_{\nu_1 ... \nu_{n_2-1} \la} \, v^{3 \ \be}_{\rho_1 ... \rho_{n_3-1} \tau} \, \langle \, \psi^\si(z_1) \, S_{\al}(z_2) \, S_\be(z_3) \, \rangle \biggr. \notag \\
& \ \ \ \ \ \ \ \ \ \ \biggl.   \langle \, i\pa X^{\mu_1} \, ... \, i \pa X^{\mu_{n_1}}(z_1) \, i\pa X^{\nu_1} \, ... \, i \pa X^{\nu_{n_2-1}} \, i \pa X^\la(z_2) \, i\pa X^{\rho_1} \, ... \, i \pa X^{\rho_{n_3-1}} \, i \pa X^{\tau}(z_3) \, \prod_{j=1}^3 e^{ik_j X(z_j)} \, \rangle
\biggr. \notag \\
& \ \ \ \ \ \ \ \  \biggl. + \ 2 \ap \, v^{2 \ \al}_{\nu_1 ... \nu_{n_2-1} \la} \, \rhob^{3 \ \tau}_{\rho_1 ... \rho_{n_3-1}  \dbe} \, \langle \, \psi^\si(z_1) \, S_{\al}(z_2) \, K_\tau^{\dbe}(z_3) \, \rangle \biggr. \notag \\
& \ \ \ \ \ \ \ \ \ \ \biggl.  \langle \, i\pa X^{\mu_1} \, ... \, i \pa X^{\mu_{n_1}}(z_1) \, i\pa X^{\nu_1} \, ... \, i \pa X^{\nu_{n_2-1}} \, i \pa X^\la(z_2) \, i\pa X^{\rho_1} \, ... \, i \pa X^{\rho_{n_3-1}}(z_3) \,  \prod_{j=1}^3 e^{ik_j X(z_j)} \, \rangle
\biggr. \notag \\
& \ \ \ \ \ \ \ \ \biggl. + \ 2 \ap \, \rhob^{2 \ \la}_{\nu_1 ... \nu_{n_2-1} \dal} \, v^{3 \ \be}_{\rho_1 ... \rho_{n_3-1}  \tau} \, \langle \, \psi^\si(z_1) \, K_{\la}^{\dal}(z_2) \, S_\be(z_3) \, \rangle \biggr. \notag \\
& \ \ \ \ \ \ \ \ \ \ \biggl. \langle \, i\pa X^{\mu_1} \, ... \, i \pa X^{\mu_{n_1}}(z_1) \, i\pa X^{\nu_1} \, ... \, i \pa X^{\nu_{n_2-1}} (z_2) \, i\pa X^{\rho_1} \, ... \, i \pa X^{\rho_{n_3-1}} \, i \pa X^{\tau}(z_3) \, \prod_{j=1}^3 e^{ik_j X(z_j)} \, \rangle
\biggr. \notag \\
& \ \ \ \ \ \ \ \ \biggl. + \ 4 \ap^2 \, \rhob^{2 \ \la}_{\nu_1 ... \nu_{n_2-1} \dal} \, \rhob^{3 \ \tau}_{\rho_1 ... \rho_{n_3-1}  \dbe} \, \langle \, \psi^\si(z_1) \, K^{\dal}_\la(z_2) \, K_\tau^{\dbe}(z_3) \, \rangle \biggr. \notag \\
& \ \ \ \ \ \ \ \ \ \ \biggl.   \langle \, i\pa X^{\mu_1} \, ... \, i \pa X^{\mu_{n_1}}(z_1) \, i\pa X^{\nu_1} \, ... \, i \pa X^{\nu_{n_2-1}} (z_2) \, i\pa X^{\rho_1} \, ... \, i \pa X^{\rho_{n_3-1}}(z_3) \,  \prod_{j=1}^3 e^{ik_j X(z_j)} \, \rangle\,
\biggr\}
\label{f1}
\end{align}
The superghost correlator in this amplitude reads $\langle e^{-\phi(z_1)} e^{-\phi(z_2)/2} e^{-\phi(z_3)/2} \rangle = (z_{12} z_{13})^{-1/2} z_{23}^{-1/4}$, the fractional power of $z_{ij}$ are completed to integers by the spin field correlation functions given in appendix \ref{appB}. After appropriately labelling summation indices $i,j,k$ and eliminating the $\rhob^i$ via Dirac equation, the vertex simplifies to
\begin{align}
{\cal A}_{(s_1,n_2+\frac{1}{2},n_3+\frac{1}{2})}& \ \ = \ \ \frac{\sqrt{2\ap}^{n_1+n_2+n_3}}{\sqrt{2}} \; n_1 ! \,\sum_{i,j,k \in {\cal I}_f} \frac{(n_2-1)! \, (n_3 - 1)! \,   (2\ap)^{-i-j-k} \, \de^{i}_{12} \, \de^j_{13} \, \de^k_{23}  }{i! \, j! \, k! \, (s_1 - i - j)! \, (n_2 - i - k)! \, (n_3 - j - k)!} \notag \\
&\times \ \biggl\{ \, n_2 \, n_3 \, (s_1 - i - j) \, (\phi_\mu \cdot k_2^{n_1 - i - j}) \, (k_3^{n_2 - i - k} \cdot v_2^{\al}) \, \ga^\mu_{\al \be} \, (v_3^{\be} \cdot k_1^{n_3 - j - k}) \, \biggr. \notag \\
& \ \ \ \ \ \biggl. + \ j \, n_2 \, 2\ap \, (\phi \cdot k_2^{n_1+1 - i - j}) \, (k_3^{n_2 - i - k} \cdot v_2^{\al})  \not \! k_{3\al \be} \, (v_3^{\be} \cdot k_1^{n_3 - j - k}) \, \biggr. \notag \\
& \ \ \ \ \ \biggl. - \ i \, n_3 \, 2\ap \, (\phi \cdot k_2^{n_1 +1 - i - j}) \, (k_3^{n_2 - i - k} \cdot v_2^{\al})  \not \! k_{2\al \be} \, (v_3^{\be} \cdot k_1^{n_3 - j - k}) \, \biggr. \notag \\
& \ \ \ \ \ \biggl. - \ \ap \, k \, (s_1 - i - j ) \, (\phi_\mu \cdot k_2^{n_1 - i - j}) \, (k_3^{n_2 - i - k} \cdot v_2^{\al}) \, (\not \! k_2 \, \gab^\mu \! \not \! k_3)_{\al \be} \, (v_3^{\be} \cdot k_1^{n_3 - j - k}) \, \biggr\}
\label{f2}
\end{align}
with summation range
\begin{align}
{\cal I}_f \ \ := \ \ \Bigl\{  \, i,j,k \in \NN_0 \, : \ s_1 - i - j \geq 0 \ , \ \ \ n_2 - i-k \geq 0 \ , \ \ \ n_3 - j - k \geq 0 \, \Bigr\} \ .
\label{f3}
\end{align}
Note that we have absorbed the occurring charge conjugation matrices into the gamma matrices, i.e. $\ga^\mu_{\al \be} \equiv \ga^\mu_{\al \dbe} C^{\dbe} {}_{\be}$. They are symmetric in $D=10$ dimensions $\ga^\mu_{\al \be} = \ga^\mu_{(\al \be)}$.

\medskip
Let us explicitly display the electromagnetic coupling with $\xi_\mu \equiv \phi_\mu$ to higher spin fermions, firstly at coinciding mass level $n_2 = n_3 \equiv n$:
\begin{align}
{\cal A}&_{(1,n+\frac{1}{2},n+\frac{1}{2})} \ \ = \ \ \frac{(2\ap)^n}{\sqrt{2}} \; \bigl[ \, (n-1)! \, \bigr]^2 \sum_{k=0}^n \frac{ (2\ap)^{-k} \, \de_{23}^k}{ k! \, \bigl[ \, (n-k)! \, \bigr]^2} \notag \\
&\ \ \ \ \ \times \ \biggl\{ \, n^2 \, (k_3^{n - k} \cdot v_2^{\al}) \not \! \xi_{\al \be} \, (v_3^{\be} \cdot k_1^{n- k}) \ - \ \ap \, k \,  (k_3^{n - k} \cdot v_2^{\al}) \, (\not \! k_{2} \! \not \! \xi \! \not \! k_3)_{\al \be} \, (v_3^{\be} \cdot k_1^{n- k}) \, \biggr. \notag \\
&\ \ \ \ \ \ \ \ \ \ \biggl. + \ n \, (n-k)\, \xi^\mu \, \Bigl[ \, (k_3^{n - k} \cdot v_2^{\al}) \not \! k_{3 \al \be} \, (v_{3\mu}^{\be} \cdot k_1^{n- k-1})  \ - \ (k_3^{n - k-1} \cdot v_{2\mu}^{\al}) \not \! k_{2 \al \be} \, (v_{3}^{\be} \cdot k_1^{n- k})  \, \Bigr]
 \, \biggr\}
\label{f4}
\end{align}
For different fermion spins $s_2 < s_3$, on the other hand,
\begin{align}
{\cal A}&_{(1,n_2+\frac{1}{2},n_3+\frac{1}{2})} \ \ = \ \ \frac{\sqrt{2\ap}^{n_2+n_3}}{\sqrt{2}} \; (n_2-1)! \, (n_3 - 1)! \sum_{k=0}^{n_2} \frac{ (2\ap)^{-k} \, \de_{23}^k}{ k! \, (n_2 -k)! \, (n_3 - k)!} \notag \\
&\ \ \ \ \ \times \ \biggl\{ \, n_2 \, n_3 \, (k_3^{n_2 - k} \cdot v_2^{\al}) \not \! \xi_{\al \be} \, (v_3^{\be} \cdot k_1^{n_3- k}) \ - \ \ap \, k \,  (k_3^{n_2 - k} \cdot v_2^{\al}) (\not \! k_{2} \not \! \xi \not \! k_3)_{\al \be} \, (v_3^{\be} \cdot k_1^{n_3- k}) \, \biggr. \notag \\
&\ \ \ \ \ \ \ \ \ \ \biggl. + \ n_2 \, (n_3-k) \,  (k_3^{n_2 - k} \cdot v_2^{\al}) \not \! k_{3 \al \be} \, (v_{3\mu}^{\be} \cdot k_1^{n_3- k-1}) \, \xi^\mu \biggr. \notag \\
&\ \ \ \ \ \ \ \ \ \ \biggl. - \ n_3 \, (n_2-k) \, \xi^\mu \, (k_3^{n_2 - k-1} \cdot v_{2\mu}^{\al}) \not \! k_{2 \al \be} \, (v_{3}^{\be} \cdot k_1^{n_3- k}) 
 \, \biggr\}
\label{f5}
\end{align}
Another interesting special case is one massless Weyl fermion of spin $s_2 = \frac{1}{2}$ whose wave function $u^\al$ satisfies the massless Dirac equation $u^\al \! \not \! k_{2\al \dbe} = 0$:
\begin{align}
{\cal A}&_{(s_1,\frac{1}{2},n_3+\frac{1}{2})} \ \ = \ \ \frac{\sqrt{2\ap}^{n_1+n_3}}{\sqrt{2}} \, n_1! \, (n_3-1)! \sum_{j=0}^{\te{min}(n_1+1,n_3)} \frac{(2\ap)^{-j} \, \de^j_{13}   }{  j! \, (s_1 - j)! \, (n_3 - j )!} \notag \\
&\times \ \biggl\{ \, n_3 \, (s_1 - j) \, (\phi_\mu \cdot k_2^{n_1 - j}) \, u^\al \, \ga^\mu_{\al \be} \, (v_3^{\be} \cdot k_1^{n_3 - j }) \ + \ j \, 2\ap \, (\phi \cdot k_2^{n_1+1 - j}) \, u^\al \! \not \! k_{3\al \be} \, (v_3^{\be} \cdot k_1^{n_3 - j }) \, \biggr\}
\label{f6}
\end{align}

\subsection{Dirac spinors and chiral interactions}
\label{sec:Dirac}

The fermion interactions from superstring theory are naturally chiral since the vertex operator (\ref{3,1}) clearly distinguishes between left handed and right handed tensor spinors $v$ and $\rhob$. To work out the degree of asymmetry in the left- and right handed couplings, it is convenient to use Dirac spinor notation, see appendix \ref{appC} for our conventions.

\medskip
Let us express our three point vertices from the previous subsection in terms of $\Ga^\mu, \Ga_{11}$ and $\Psi^A = \left( v^\al, \, \bar w_{\dal} \right)^t$ with $\bar w_{\dal} = \sqrt{ \frac{\ap}{ n} } (D-2) \rhob_{\dal}$. The most general case (\ref{f2}) with spins $s_1 = n_1+1$ and $s_{2,3} = n_{2,3} + \frac{1}{2}$ reads as follows in Dirac spinor language,
\begin{align}
{\cal A}_{(s_1,n_2+\frac{1}{2},n_3+\frac{1}{2})}& \ \ = \ \ \frac{\sqrt{2\ap}^{n_1+n_2+n_3}}{\sqrt{2}} \; n_1 ! \,  \sum_{i,j,k \in {\cal I}_f} \frac{(n_2-1)! \, (n_3 - 1)! \, (2\ap)^{-i-j-k} \, \de^{i}_{12} \, \de^j_{13} \, \de^k_{23}  }{i! \, j! \, k! \, (s_1 - i - j)! \, (n_2 - i - k)! \, (n_3 - j - k)!} \notag \\
\times \ &\biggl\{ \, \frac{ \sqrt{n_2 \, n_3}}{2} \; (s_1 - i - j) \, (\phi_\mu \cdot k_2^{n_1 - i - j}) \, (k_3^{n_2 - i - k} \cdot \Psi_2^{A}) \biggr. \notag \\
& \biggl. \ \ \ \ \ \ \ \ \ \ \ \ \ \ \ \bigl[ \,  (\sqrt{n_2 \, n_3} \, + \, k) \, \Ga^\mu \ + \ (\sqrt{n_2 \, n_3} \, - \, k)  \, \Ga^\mu \, \Ga_{11} \, \bigr]_{AB} \, (\Psi_3^{B} \cdot k_1^{n_3 - j - k}) \, \biggr. \notag \\
&   \biggl. + \ \sqrt{n_2 \, n_3 \, \ap}  \, (\phi \cdot k_2^{n_1+1 - i - j}) \, (k_3^{n_2 - i - k} \cdot \Psi_2^{A}) \,  \biggr. \notag \\
& \biggl. \ \ \ \ \ \ \ \ \ \ \ \ \ \ \ \bigl[ \, 
 (i \sqrt{n_3} \, + \, j \, \sqrt{n_2}) \, \mathds{1} \ + \ \, (  i \sqrt{n_3} \, - \, j \, \sqrt{n_2}) \, \Ga_{11} \, \bigr]_{AB} \, (\Psi_3^{B} \cdot k_1^{n_3 - j - k}) \, \biggr\} \ ,
\label{Dirac5}
\end{align}
see (\ref{Dirac4}) for the details of the conversion. Note that equal fermion masses $n_2 = n_3 = n$ give rise to some cancellations in the axial $\Ga_{11}$ terms: Firstly, the lowest derivative contribution $\Psi^A_{2 \mu_1 ... \mu_n} \Ga^\mu_{AB} \Psi^{B\mu_1 ... \mu_n}_3$ at $(i,j,k) = (0,0,n)$ is nonchiral and can be obtained from a covariant derivative $D_\la = \pa_\la + A_\la$ in the effective Lagrangian $\Psi_{\mu_1 ... \mu_n} (\Ga^\la D_\la + m) \Psi^{\mu_1 ... \mu_n}$ if the boson is a massless gauge field $n_1 = 0$. Secondly, the $\Psi^A_2 (\Ga_{11})_{AB} \Psi^B_3$ terms vanish if the polarization tensor $\phi$ of the boson has the same number of contractions $i=j$ with $\Psi_2$ and $\Psi_3$.

\medskip
Let us explicitly display the electromagnetic coupling to fermions of the same mass level $n$:
\begin{align}
{\cal A}&_{(1,n+\frac{1}{2},n+\frac{1}{2})} \ \ = \ \ \frac{(2\ap)^n}{\sqrt{2}} \; n! \, (n-1)!  \, \sum_{k=0}^n \frac{ (2\ap)^{-k} \, \de_{23}^k}{ k! \, \bigl[ \, (n-k)! \, \bigr]^2} \notag \\
&\ \ \ \ \ \times \ \biggl\{ \, \frac{1}{2} \; (k_3^{n - k} \cdot \Psi_2^{A}) \, \bigl[ \, (n \, + \, k) \! \not \! \xi \ + \ (n\, - \, k) \! \not \! \xi \, \Ga_{11} \, \bigr]_{AB} \, (\Psi_3^{B} \cdot k_1^{n-k}) \biggr. \notag \\
&\ \ \ \ \ \ \ \ \ \ \biggl. + \ \frac{ \sqrt{n} \, (n-k) }{\sqrt{\ap}} \; (k_3^{n-k-1} \cdot \Psi_{2\mu}^{A}) \, \bigl[ \, \xi^{[\mu} \, k_1^{\nu]} \, \mathds{1} \ + \ \xi^{(\mu} \, k_1^{\nu)} \, \Ga_{11} \, \bigr]_{AB} \, (\Psi_{3\nu}^{B} \cdot k_1^{n- k-1})  \, \Bigr]
 \, \biggr\}
\label{Dirac6}
\end{align}
The highest $k$ contributions from both lines agree with equation (36) of \cite{ferm}\footnote{A typo $k_2^\al \leftrightarrow A_2^\al$ has sneaked into into the second line of equation (36), at least in the arXiv version of \cite{ferm}.}.

\section{Four point amplitudes with one massive state}
\label{sec:4pt1}

Four point superstring amplitudes have a much richer structure than three point couplings. Of course, the number of different tensor structures increases drastically, but the more important point is that the overall worldsheet correlators now get a non-trivial $z$ dependence which gives rise to Euler Beta functions of the Mandelstam variables when integrated over the worldsheet boundary. 

\medskip
To follow the notation of \cite{lhc1,lhc2,lhc3}, let us introduce dimensionless Mandelstam invariants
\beq
s \ \ := \ \ \ap \, (k_1 + k_2)^2 \co t \ \ := \ \ \ap \, (k_1+ k_3)^2 \co u \ \ := \ \ \ap \, (k_1 + k_4)^2
\label{2,10}
\eeq
subject to momentum conservation $s+t+u = -\ap (m_1^2+m_2^2+m_3^2+m_4^2)$ as well as the string formfactor $V_t$ in terms of the Euler Beta function $B(s,u)$:
\begin{align}
B(s,u) \ \ :=& \ \ \int \dd x \ x^{s-1} \, (1-x)^{u-1} \eq \frac{ \Ga(s) \, \Ga(u)}{\Ga(s+u)} 
\label{2,10a} \\
V_t \ \ :=& \ \ \frac{s \, u}{s+u} \; B(s,u) \eq \frac{ \Ga(s+1) \, \Ga(u+1)}{\Ga(s+u+1)} \ .
\label{2,11}
\end{align}
This definition of $V_t$ helps a lot to keep track of kinematic poles since it has a regular expansion
\beq
V_t \eq 1 \ - \, \zeta(2) \, su \ - \, \zeta(3) \, su \, (s+u) \ + \ {\cal O}(\ap^4) 
\label{2,12}
\eeq
and can be regarded as a stringy formfactor.

\subsection{Relating different color orderings}

Four point amplitudes are the first instance where distinct color ordered amplitudes exists which differ in more than just their sign like for three point. More precisely, the six cyclically inequivalent orderings of total mass level $N= \sum_i n_i$ states can be grouped into three pairs each of which is connected by worldsheet parity
\beq
{\cal A}_{(s_1,s_2,s_3,s_4)} \eq (-1)^{N} \, {\cal A}_{(s_4,s_3,s_2,s_1)} \ .
\label{sub1}
\eeq
After placing the $(-1)^N$ sign into the Chan Paton traces, we can organize the full amplitude into three contribution with
\beq
{\cal A}_t(s_i) \ \ := \ \ {\cal A}_{(s_1,s_2,s_3,s_4)} \co {\cal A}_s(s_i) \ \ := \ \ {\cal A}_{(s_2,s_3,s_1,s_4)} \co {\cal A}_u(s_i) \ \ := \ \ {\cal A}_{(s_3,s_1,s_2,s_4)}
\label{sub1a}
\eeq
which single out one of the $s$, $t$ or $u$-channel each:
\begin{align}
{\cal A}^{\te{full}}_{(s_1,s_2,s_3,s_4)} \ \ &= \ \ \sum_{\rho \in S_3} \te{Tr} \bigl\{ T^{a_{\rho(1)}} \, T^{a_{\rho(2)}} \, T^{a_{\rho(3)}} \, T^{a_4} \bigr\} \, {\cal A}_{(s_{\rho(1)},s_{\rho(2)},s_{\rho(3)},s_4)} \notag \\
&= \ \ \te{Tr} \bigl\{ T^{a_{1}} \, T^{a_{2}} \, T^{a_{3}} \, T^{a_4} \ + \ (-1)^{N} \, T^{a_{4}} \, T^{a_{3}} \, T^{a_{2}} \, T^{a_1} \bigr\} \, {\cal A}_t(s_i) \notag \\
& \ \ \ \ + \ \te{Tr} \bigl\{ T^{a_{2}} \, T^{a_{3}} \, T^{a_{1}} \, T^{a_4} \ + \ (-1)^{N} \, T^{a_{4}} \, T^{a_{1}} \, T^{a_{3}} \, T^{a_2} \bigr\} \, {\cal A}_s(s_i) \notag \\
& \ \ \ \ + \ \te{Tr} \bigl\{ T^{a_{3}} \, T^{a_{1}} \, T^{a_{2}} \, T^{a_4} \ + \ (-1)^{N} \, T^{a_{4}} \, T^{a_{2}} \, T^{a_{1}} \, T^{a_3} \bigr\} \, {\cal A}_u(s_i)
\label{sub2}
\end{align}
The three color stripped amplitudes in (\ref{sub2}) are still not independent, a worldsheet monodromy analysis \cite{mon1,mon2} yields relations
\beq
\sin (\pi s) \, {\cal A}_u \ \ = \ \ \sin (\pi u) \, {\cal A}_s \co
\sin (\pi u) \, {\cal A}_t \ \ = \ \ \sin (\pi t) \, {\cal A}_u \co
\sin (\pi t) \, {\cal A}_{s} \ \ = \ \ \sin (\pi s) \, {\cal A}_{t} \label{sub3} 
\eeq
between the three channel specific subamplitudes ${\cal A}_t, {\cal A}_s, {\cal A}_u$. Effectively, there is only one color ordered amplitude left to determine from the scratch.

\medskip
Momentum conservation $s+t+u=-N \equiv - \al' \sum_i m_i^2$ admits to rewrite the $t$ channel formfactor as $V_t = \frac{ \Ga(s+1) \Ga(u+1) }{\Ga(1-t-N)}$, then relations (\ref{sub3}) together with the Euler reflection formula $\Ga(1-z) \Ga(z) = \frac{\pi}{\sin(\pi z)}$ implies that subamplitudes can be factorized into a channel dependent kinematic part and an universal piece ${\cal A}_0$ which is common to all the three contributions in (\ref{sub2}):
\beq
{\cal A}_{t} \eq \frac{V_t}{\prod_{k=1}^{N-1}(t+k)} \; {\cal A}_0 \co {\cal A}_{s} \eq \frac{V_s}{\prod_{k=1}^{N-1}(s+k)} \; {\cal A}_0 \co {\cal A}_{u} \eq \frac{V_u}{\prod_{k=1}^{N-1}(u+k)} \; {\cal A}_0
\label{sub4}
\eeq
The massless case $N=0$ is somehow exceptional with
\beq
{\cal A}_{t} \Bigl. \, \Bigr|_{N=0} \eq \frac{V_t}{su} \; {\cal A}_0 \Bigl. \, \Bigr|_{N=0} \co {\cal A}_{s} \Bigl. \, \Bigr|_{N=0} \eq \frac{V_s}{tu} \; {\cal A}_0 \Bigl. \, \Bigr|_{N=0} \co {\cal A}_{u} \Bigl. \, \Bigr|_{N=0} \eq \frac{V_u}{st} \; {\cal A}_0 \Bigl. \, \Bigr|_{N=0} \ .
\label{sub5}
\eeq
In case of identical external states $s_i = s_j$, the universal part ${\cal A}_0$ must certainly be (anti-)symmetric under exchange of labels $i \leftrightarrow j$. Our guiding principle in presenting the following four point amplitudes will be making this symmetry manifest.

\medskip
The bottom line of these arguments is the following simple formula which translates the color ordered subamplitudes ${\cal A}_{(s_1,s_2,s_3,s_4)}$ we are about to compute into full color dressed four point amplitudes:
\begin{align}
{\cal A}^{\te{full}}_{(s_1,s_2,s_3,s_4)} \ \ = \ \ &{\cal A}_{(s_1,s_2,s_3,s_4)} \,\biggl( \, \te{Tr} \bigl\{ T^{a_{1}} \, T^{a_{2}} \, T^{a_{3}} \, T^{a_4} \ + \ (-1)^{N} \, T^{a_{4}} \, T^{a_{3}} \, T^{a_{2}} \, T^{a_1} \bigr\} \biggr. \notag \\
\biggl. &\ \ \ \ \ + \ \frac{ V_s}{V_t} \, \prod_{k=1}^{n-1} \frac{t+k}{s+k} \; \te{Tr} \bigl\{ T^{a_{2}} \, T^{a_{3}} \, T^{a_{1}} \, T^{a_4} \ + \ (-1)^{N} \, T^{a_{4}} \, T^{a_{1}} \, T^{a_{3}} \, T^{a_2} \bigr\} \biggr. \notag \\
\biggl. &\ \ \ \ \ + \ \frac{ V_u}{V_t} \, \prod_{k=1}^{n-1} \frac{t+k}{u+k} \; \te{Tr} \bigl\{ T^{a_{3}} \, T^{a_{1}} \, T^{a_{2}} \, T^{a_4} \ + \ (-1)^{N} \, T^{a_{4}} \, T^{a_{2}} \, T^{a_{1}} \, T^{a_3} \bigr\} \, \biggr)
\label{sub6}
\end{align}

\subsection{Three massless bosons, one higher spin boson} 

Given this notation, we compute the four point coupling of spin $s = n+1$ to three gluons:
\begin{align}
{\cal A}&_{(1,1,1,s)} \ \ = \ \ \langle \, cV^{(0)}(\xi^1,k_1) \, \int \dd z_2 \, V^{(-1)} (\xi^2,k_2) \, cV^{(0)} (\xi^3 , k_3) \, c V_n^{(-1)}(\phi, k_4) \, \rangle \notag \\
&= \ \ \sqrt{2\ap}^{-n-2} \, \int \dd z_2 \ \langle \,c(z_1) \, c(z_3) \, c(z_4) \, \rangle \, \langle \, e^{-\phi(z_2) } \, e^{-\phi(z_4)} \, \rangle \, \phi^4_{\tau_1 ... \tau_n \xi} \notag \\
& \ \ \ \times \, \biggl\{ \, \xi^1_\mu \, \xi^3_\la \, \langle \, i \pa X^\mu(z_1) \, i\pa X^\la(z_3) \, i \pa X^{\tau_1} \, ... \, i \pa X^{\tau_n}(z_4) \prod_{j=1}^4 e^{ik_j X(z_j)} \, \rangle \, \xi_\ka^2 \, \langle \, \psi^\ka(z_2) \, \psi^\xi(z_4) \, \rangle \biggr. \notag \\
& \ \ \ \ \biggl. + \, 4 \ap^2 \, \langle \, i \pa X^{\tau_1} \, ... \, i \pa X^{\tau_n}(z_4) \prod_{j=1}^4 e^{ik_j X(z_j)} \, \rangle \, \xi_\mu^1 \, k_\nu^1 \, \xi_\la^3 \, k_\rho^3 \, \xi_\ka^2 \, \langle \, \psi^{\mu} \, \psi^\nu (z_1) \, \psi^\ka(z_2) \, \psi^\la \, \psi^\rho(z_3) \, \psi^\xi(z_4) \, \rangle \biggr. \notag \\
& \ \ \ \ \biggl. - \, 2 \ap \, \xi_\la ^3 \,  \langle \, i \pa X^\la (z_3) \, i \pa X^{\tau_1} \, ... \, i \pa X^{\tau_n}(z_4) \prod_{j=1}^4 e^{ik_j X(z_j)} \, \rangle \, \xi^1_\mu \, k_\nu^1 \, \xi_\ka^2 \, \langle \, \psi^\mu \, \psi^\nu(z_1) \, \psi^{\ka}(z_2) \, \psi^\xi(z_4) \, \rangle \biggr. \notag \\
& \ \ \ \ \biggl. - \, 2 \ap \, \xi_\mu ^1 \,  \langle \, i \pa X^\mu (z_1) \, i \pa X^{\tau_1} \, ... \, i \pa X^{\tau_n}(z_4) \prod_{j=1}^4 e^{ik_j X(z_j)} \, \rangle \, \xi^3_\la \, k_\rho^3 \, \xi_\ka^2 \, \langle \,  \psi^{\ka}(z_2) \, \psi^\la \, \psi^\rho(z_3) \, \psi^\xi(z_4) \, \rangle \, \biggr\}
\label{2,13}
\end{align}
All the correlation functions can be taken from appendix \ref{appA}. It is convenient to map the worldsheet boundary to the reals axis and to fix $(z_1,z_3,z_4) = (0,1,\infty)$ such that $z_2$ is integrated over the $[0,1]$ interval to ensure cyclic ordering $(1,2,3,4)$ of the insertion points. This gives rise to Euler Beta functions (\ref{2,10}):
\begin{align}
{\cal A}&_{(1,1,1,n+1)} \ \ = \ \ \sqrt{2\ap}^{n+2} \, (-1)^{n-1} \notag \\
& \biggl\{ \, \frac{n \, (n-1)}{4\ap^2} \; \xi^1_\mu \, \xi^2_\nu \, \xi^3_\la \sum_{p=0}^{n-2} \left( \begin{smallmatrix} n-2 \\ p \end{smallmatrix} \right) \, \bigl[ \phi^{\mu \nu \la} \cdot k_1^p \, (-k_3)^{n-2-p} \bigr] \, B(s+1+p,u+n-1-p) \biggr. \notag \\
\biggl.& \ + \ \frac{1-t}{2\ap} \; (\xi^1 \, \xi^3) \, \xi^\mu_2 \sum_{p=0}^n 
\left( \begin{smallmatrix} n \\ p \end{smallmatrix} \right) \, \bigl[ \phi^{\mu \nu \la} \cdot k_1^p \, (-k_3)^{n-p} \bigr] \, B(s+1+p,u+n+1-p) \biggr.
 \notag \\
 \biggl.& \ + \ \Bigl[ \, (\xi^1 \, k_2) \, (\xi^3 \, k_1) \, \xi^2_\mu \, k^2_\nu  \ - \ (\xi^1 \, \xi^2) \, (\xi^3 \, k_1) \, k^2_\mu \, k^2_\nu  \ + \ (\xi^3 \, \xi^2) \, (\xi^1 \, k_3) \, k^2_\mu \, k^2_\nu \ - \ (\xi^3 \, k_2) \, (\xi^1 \, k_3) \, \xi^2_\mu \, k^2_\nu  \Bigr. \biggr. \notag \\
 \biggl. \Bigl. & \ \ \ \ \ \ \ \ \ \ + \ (\xi^2 \, k_4) \, (\xi^3 \, k_1) \, \xi^1_\mu \, k_\nu^2 \ - \ (\xi^2 \, k_4) \, (\xi^1 \, k_3) \, \xi^3_\mu \, k_\nu^2 \ + \ (\xi^1 \, \xi^3) \, (\xi^2 \, k_3) \, k_\mu^1 \, k_\nu^2 \ - \ (\xi^1 \, \xi^3) \, (\xi^2 \, k_1) \, k_\mu^3 \, k_\nu^2
  \Bigr. \biggr. \notag \\
 \biggl. \Bigl. & \ \ \ \ \ \ \ \ \ \ + \ \frac{t}{2\ap} \; (\xi^2 \, \xi^3) \, \xi^1_\mu \, k^2_\nu \ - \ \frac{t}{2\ap} \; (\xi^1 \, \xi^2) \, \xi^3_\mu \, k^2_\nu \ + \ \frac{n}{2\ap} \; (\xi^1 \, k_3) \, \xi^2_\mu \, \xi^3_\nu \ - \ \frac{n}{2\ap} \; (\xi^3 \, k_1) \, \xi^1_\mu \, \xi^2_\nu \, \Bigr] \notag \\
 \biggl. &  \ \ \ \ \ \ \ \ \ \ \ \ \ \sum_{p=0}^{n-1} 
\left( \begin{smallmatrix} n-1 \\ p \end{smallmatrix} \right) \, \bigl[ \phi^{\mu \nu} \cdot k_1^p \, (-k_3)^{n-1-p} \bigr] \, B(s+1+p,u+n-p) \biggr.
 \notag \\
   \biggl.& \ + \ \Bigl[ \, (\xi^1 \, k_3) \, (\xi^2 \, k_1) \, \xi^3_\mu \, k^3_\nu  \ - \ (\xi^1 \, \xi^3) \, (\xi^2 \, k_1) \, k^3_\mu \, k^3_\nu \ + \ (\xi^2 \, \xi^3) \, (\xi^1 \, k_2) \, k^3_\mu \, k^3_\nu \ - \  (\xi^2 \, k_3) \, (\xi^1 \, k_2) \, \xi^3_\mu \, k^3_\nu \Bigr. \biggr. \notag \\
   \biggl. \Bigl. & \ \ \ \ \ \ \ \ \ \ + \ (\xi^3 \, k_4) \, (\xi^2 \, k_1) \, \xi^1_\mu \, k_\nu^3 \ - \ (\xi^3 \, k_4) \, (\xi^1 \, k_2) \, \xi^2_\mu \, k_\nu^3 \ + \ (\xi^1 \, \xi^2) \, (\xi^3 \, k_2) \, k_\mu^1 \, k_\nu^3 \ - \ (\xi^1 \, \xi^2) \, (\xi^3 \, k_1) \, k_\mu^2 \, k_\nu^3
  \Bigr. \biggr. \notag \\
   \biggl. \Bigl. & \ \ \ \ \ \ \ \ \ \ + \ \frac{n}{2\ap} \; (\xi^1 \, k_2) \, \xi^2_\mu \, \xi^3_\nu \ - \ \frac{n}{2\ap} \; (\xi^2 \, k_1) \, \xi^1_\mu \, \xi^3_\nu \ + \ \frac{n}{2\ap} \; (\xi^1 \, \xi^2) \, \xi^3_\mu \, k^1_\nu \ - \ \frac{t}{2\ap} \; (\xi^1 \, \xi^2) \, \xi^3_\mu \, k^3_\nu \, \Bigr] \notag \\
  \biggl. &  \ \ \ \ \ \ \ \ \ \ \ \ \ \sum_{p=0}^{n-1} 
\left( \begin{smallmatrix} n-1 \\ p \end{smallmatrix} \right) \, \bigl[ \phi^{\mu \nu} \cdot k_2^p \, k_1^{n-1-p} \bigr] \, B(s,u+1+p) \biggr.
 \notag \\
   \biggl.& \ + \ \Bigl[ \, (\xi^2 \, k_1) \, (\xi^3 \, k_2) \, \xi^1_\mu \, k^1_\nu  \ - \ (\xi^1 \, \xi^2) \, (\xi^3 \, k_2) \, k^1_\mu \, k^1_\nu \ + \ (\xi^1 \, \xi^3) \, (\xi^2 \, k_3) \, k^1_\mu \, k^1_\nu \ - \  (\xi^3 \, k_1) \, (\xi^2 \, k_3) \, \xi^1_\mu \, k^1_\nu \Bigr. \biggr. \notag \\
 \biggl. \Bigl. & \ \ \ \ \ \ \ \ \ \ + \ (\xi^1 \, k_4) \, (\xi^3 \, k_2) \, \xi^2_\mu \, k_\nu^1 \ - (\xi^1 \, k_4) \, (\xi^2 \, k_3) \, \xi^3_\mu \, k_\nu^1 \ + \ (\xi^2 \, \xi^3) \, (\xi^1 \, k_3) \, k_\mu^1 \, k_\nu^2 \ - \ (\xi^2 \, \xi^3) \, (\xi^1 \, k_2) \, k_\mu^1 \, k_\nu^3
  \Bigr. \biggr. \notag \\
 \biggl. \Bigl. & \ \ \ \ \ \ \ \ \ \ + \ \frac{n}{2\ap} \; (\xi^2 \, k_3) \, \xi^1_\mu \, \xi^3_\nu \ - \ \frac{n}{2\ap} \; (\xi^3 \, k_2) \, \xi^1_\mu \, \xi^2_\nu \ - \ \frac{n}{2\ap} \; (\xi^2 \, \xi^3) \, \xi^1_\mu \, k^3_\nu \ + \ \frac{t}{2\ap} \; (\xi^2 \, \xi^3) \, \xi^1_\mu \, k^1_\nu \, \Bigr] \notag \\
  \biggl. &  \ \ \ \ \ \ \ \ \ \ \ \ \ \sum_{p=0}^{n-1} 
\left( \begin{smallmatrix} n-1 \\ p \end{smallmatrix} \right) \, \bigl[ \phi^{\mu \nu} \cdot (-k_3)^p \, (-k_2)^{n-1-p} \bigr] \, B(s+n-p,u) \, \biggr\}
\label{betas}
\end{align}
This pattern of Beta functions has already been observed in \cite{Chan:2004tb} in the context of tachyon scattering with one massive leading trajectory state on the bosonic string. One has to admit that in the presentation (\ref{betas}), it is no longer meaningful to take a massless gluon at $n=0$ for the fourth state. The nicest organization we could find for this and other four point amplitudes is guided by the $n=1$ results in \cite{lhc3}.

\medskip
Any $B(s+p,u+q)$ can be reduced to $V_t$ by factoring out appropriate Pochhammer symbols
\begin{align}
(x)_n \ \ &:= \ \ \prod_{j=0}^{n-1} (x+j) \eq x \, (x+1) \, ... \, (x+n-1)
\label{2,14} \\
\ga(x,n) \ \ &:= \ \ \frac{(-x)_n}{n!} \eq \frac{1}{n!} \; \prod_{j=0}^{n-1} (-x+j) \ .
\label{2,15}
\end{align}
In agreement with (\ref{sub4}), the result contains a prefactor $V_t \prod_{k=1}^{n-1} (t+k)^{-1}$ which is specific to the ordering $(1,2,3,4)$ of the external legs and a cyclically symmetric kinematic factor ${\cal A}_0$ under rearrangement $(1,2,3) \mapsto (2,3,1)$ of the three gluons (although this symmetry is not immediately obvious in the term $\sim \, \xi^1_\mu  \, \xi^2_\nu \,\xi^3_\la \, \phi^{\mu \nu \la}$). Moreover, exchange of two massless states $\xi^i \leftrightarrow \xi^j$ yields the $n$ dependent sign $(-)^n$ in this color stripped amplitude to compensate for the $(-)^n$ in the color factor and preserve the bosonic statistics of gluons.
\begin{align}
&{\cal A}_{(1,1,1,n+1)} \ \ = \ \  \frac{\sqrt{2\ap}^{n+2} \, (n-1)! \, V_t}{\prod_{k=1}^{n-1} (t+k)} \notag \\
&\biggl\{ \, \frac{n}{4\ap^2} \; \sum_{p=0}^{n-2}  \ga(-s-1,p) \, \ga(-u-1,n-2-p) \, \xi^1_\mu \, \xi^2_\nu \, \xi^3_\la \, \bigl[ \phi^{\mu \nu \la} \cdot k_1^p \, (-k_3)^{n-2-p} \bigr] \biggr. \notag \\
&  \biggl. - \ \frac{n}{2\ap \, s} \; \sum_{p=0}^{n} \ga(-u-1,p) \, \ga(-t-1,n-p) \, (\xi^1 \, \xi^2) \, \xi^3_\mu \, \bigl[ \phi^\mu \cdot k_2^p \, (-k_1)^{n-p} \bigr] \biggr. \notag \\
&  \biggl. - \ \frac{n}{2\ap \, t} \; \sum_{p=0}^{n} \ga(-s-1,p) \, \ga(-u-1,n-p) \, (\xi^1 \, \xi^3) \, \xi^2_\mu \, \bigl[ \phi^\mu \cdot k_1^p \, (-k_3)^{n-p} \bigr] \biggr. \notag \\
&  \biggl. - \ \frac{n}{2\ap \, u} \; \sum_{p=0}^{n} \ga(-t-1,p) \, \ga(-s-1,n-p) \, (\xi^2 \, \xi^3) \, \xi^1_\mu \, \bigl[ \phi^\mu \cdot k_3^p \, (-k_2)^{n-p} \bigr] \biggr. \notag \\
&  \biggl. + \ \frac{1}{s} \; \sum_{p=0}^{n-1} \ga(-u-1,p) \, \ga(-t-1,n-1-p) \, \bigl[ \phi^{\mu \nu} \cdot k_2^p \, (-k_1)^{n-1-p} \bigr] \, \Bigl[\, \frac{n}{2\ap} \; (\xi^1 \, k_2) \, \xi^2_\mu \, \xi^3_\nu \ - \ \frac{n}{2\ap} \; (\xi^2 \, k_1) \, \xi^1_\mu \, \xi^3_\nu \Bigr. \biggr. \notag \\
   \biggl. \Bigl. & \ \ \ \ \ \ \ \ + \  (\xi^1 \, k_3) \, (\xi^2 \, k_1) \, \xi^3_\mu \, k^3_\nu  \ - \ (\xi^1 \, \xi^3) \, (\xi^2 \, k_1) \, k^3_\mu \, k^3_\nu \ + \ (\xi^2 \, \xi^3) \, (\xi^1 \, k_2) \, k^3_\mu \, k^3_\nu \ - \  (\xi^2 \, k_3) \, (\xi^1 \, k_2) \, \xi^3_\mu \, k^3_\nu \Bigr. \biggr. \notag \\
   \biggl. \Bigl. & \ \ \ \ \ \ \ \ + \ (\xi^3 \, k_4) \, (\xi^2 \, k_1) \, \xi^1_\mu \, k_\nu^3 \ - \ (\xi^3 \, k_4) \, (\xi^1 \, k_2) \, \xi^2_\mu \, k_\nu^3 \ + \ (\xi^1 \, \xi^2) \, (\xi^3 \, k_2) \, k_\mu^1 \, k_\nu^3 \ - \ (\xi^1 \, \xi^2) \, (\xi^3 \, k_1) \, k_\mu^2 \, k_\nu^3 \,
  \Bigr] \biggr. \notag \\
%
&  \biggl. + \ \frac{1}{t} \; \sum_{p=0}^{n-1} \ga(-s-1,p) \, \ga(-u-1,n-1-p) \, \bigl[ \phi^{\mu \nu} \cdot k_1^p \, (-k_3)^{n-1-p} \bigr] \, \Bigl[\, \frac{n}{2\ap} \; (\xi^3 \, k_1) \, \xi^1_\mu \, \xi^2_\nu \ - \ \frac{n}{2\ap} \; (\xi^1 \, k_3) \, \xi^2_\mu \, \xi^3_\nu \Bigr. \biggr. \notag \\ 
 \biggl. \Bigl. & \ \ \ \ \ \ \ \ + \ (\xi^3 \, k_2) \, (\xi^1 \, k_3) \, \xi^2_\mu \, k^2_\nu  \ - \  (\xi^3 \, \xi^2) \, (\xi^1 \, k_3) \, k^2_\mu \, k^2_\nu \ + \ (\xi^1 \, \xi^2) \, (\xi^3 \, k_1) \, k^2_\mu \, k^2_\nu \ - \  (\xi^1 \, k_2) \, (\xi^3 \, k_1) \, \xi^2_\mu \, k^2_\nu \Bigr. \biggr. \notag \\
 \biggl. \Bigl. & \ \ \ \ \ \ \ \ + \ (\xi^2 \, k_4) \, (\xi^1 \, k_3) \, \xi^3_\mu \, k_\nu^2 \ - \ (\xi^2 \, k_4) \, (\xi^3 \, k_1) \, \xi^1_\mu \, k_\nu^2 \ + \ (\xi^1 \, \xi^3) \, (\xi^2 \, k_1) \, k_\mu^3 \, k_\nu^2 \ - \ (\xi^1 \, \xi^3) \, (\xi^2 \, k_3) \, k_\mu^1 \, k_\nu^2 \,
  \Bigr] \biggr. \notag \\
 %
&  \biggl. + \ \frac{1}{u} \; \sum_{p=0}^{n-1} \ga(-t-1,p) \, \ga(-s-1,n-1-p) \, \bigl[ \phi^{\mu \nu} \cdot k_3^p \, (-k_2)^{n-1-p} \bigr] \, \Bigl[\, \frac{n}{2\ap} \; (\xi^2 \, k_3) \, \xi^1_\mu \, \xi^3_\nu \ - \ \frac{n}{2\ap} \; (\xi^3 \, k_2) \, \xi^1_\mu \, \xi^2_\nu \Bigr. \biggr. \notag \\
\biggl. \Bigl. & \ \ \ \ \ \ \ \ + \  (\xi^2 \, k_1) \, (\xi^3 \, k_2) \, \xi^1_\mu \, k^1_\nu  \ - \ (\xi^1 \, \xi^2) \, (\xi^3 \, k_2) \, k^1_\mu \, k^1_\nu \ + \ (\xi^1 \, \xi^3) \, (\xi^2 \, k_3) \, k^1_\mu \, k^1_\nu \ - \  (\xi^3 \, k_1) \, (\xi^2 \, k_3) \, \xi^1_\mu \, k^1_\nu \Bigr. \biggr. \notag \\
 \biggl. \Bigl. & \ \ \ \ \ \ \ \ + \ (\xi^1 \, k_4) \, (\xi^3 \, k_2) \, \xi^2_\mu \, k_\nu^1 \ - (\xi^1 \, k_4) \, (\xi^2 \, k_3) \, \xi^3_\mu \, k_\nu^1 \ + \ (\xi^2 \, \xi^3) \, (\xi^1 \, k_3) \, k_\mu^1 \, k_\nu^2 \ - \ (\xi^2 \, \xi^3) \, (\xi^1 \, k_2) \, k_\mu^1 \, k_\nu^3 \,
  \Bigr] \, \biggr\} 
\label{2,17}
\end{align}
Note that at $n=1$, this reproduces the result of \cite{lhc3} for decay of spin two into three gluons.

\subsection{Two massless bosons, one massless fermion, one higher spin fermion}

In this section, we compute the (color ordered) four point amplitude of two massless gauge boson with one massless Weyl fermion and a heavy higher spin fermion at level $n$:
\begin{align} 
{\cal A}&_{(1,1,\frac{1}{2},n+\frac{1}{2})} \ \ = \ \ \langle \, cV^{(-1)}(\xi^1,k_1) \, \int \dd z_2 \, V^{(0)} (\xi^2,k_2) \, cV^{(-\frac{1}{2})} (u , k_3) \, c V_n^{(-\frac{1}{2})}(v,\rhob, k_4) \, \rangle \notag \\
&= \ \ \sqrt{2\ap}^{-n-1} \, \int \dd z_2 \ \langle \,c(z_1) \, c(z_3) \, c(z_4) \, \rangle \, \langle \, e^{-\phi(z_1) } \, e^{-\phi(z_3)/2} \, e^{-\phi(z_4)/2} \, \rangle \, \xi^1_\mu \, \xi^2_\nu \, u^\al \notag \\
& \ \ \ \times \, \biggl\{ \, v^\be_{\la_1 ... \la_n} \, \langle \, i \pa X^\nu(z_2) \, i \pa X^{\la_1} \, ... \, i \pa X^{\la_n}(z_4) \prod_{j=1}^4 e^{ik_j X(z_j)} \, \rangle \, \langle \, \psi^\mu(z_1) \, S_\al(z_3) \, S_\be(z_4) \, \rangle \biggr. \notag \\
& \ \ \ \ \biggl. - \, 4 \ap^2 \, \rhob^\ka_{\la_1 ... \la_{n-1} \dbe} \, \langle \, i \pa X^{\la_1} \, ... \, i \pa X^{\la_{n-1}}(z_4) \prod_{j=1}^4 e^{ik_j X(z_j)} \, \rangle \, k_\tau^2 \, \langle \, \psi^{\mu}(z_1) \, \psi^\nu \, \psi^\tau(z_2) \, S_\al(z_3) \, K_\ka^{\dbe}(z_4) \, \rangle \biggr. \notag \\
& \ \ \ \ \biggl. - \, 2 \ap \, v_{\la_1...\la_n}^\be \,  \langle \, i \pa X^{\la_1} \, ... \, i \pa X^{\la_n}(z_4) \prod_{j=1}^4 e^{ik_j X(z_j)} \, \rangle \, k_\tau^2 \, \langle \, \psi^\mu(z_1) \, \psi^\nu \, \psi^\tau(z_2) \, S_\al(z_3) \, S_\be(z_4) \, \rangle \biggr. \notag \\
& \ \ \ \ \biggl. + \, 2 \ap \, \rhob^\ka_{\la_1...\la_{n-1}\dbe} \,  \langle \, i \pa X^\nu (z_2) \, i \pa X^{\la_1} \, ... \, i \pa X^{\la_{n-1}}(z_4) \prod_{j=1}^4 e^{ik_j X(z_j)} \, \rangle \, \langle \,  \psi^{\mu}(z_1) \, S_\al(z_3) \, K_\ka^{\dbe}(z_4) \, \rangle \, \biggr\}
\label{3,51}
\end{align}
Terms involving $\rhob$ nicely conspire and appear in groups $\sim (D-2)$ only. We can then use the Dirac equation to eliminate $\rhob$ in favor of $\not \! k_4 v$:
\begin{align}
&{\cal A}_{(1,1,\frac{1}{2},n+\frac{1}{2})} \ \ = \ \ \frac{\sqrt{2\ap}^{n+1}}{\sqrt 2} \; (-1)^{n-1} \notag \\
& \biggl\{ \, \frac{n-1}{2\ap} \; \xi^\mu_1 \, \xi^\nu_2 \, (u \! \not \! k_4)_\al \sum_{p=0}^{n-2}  \left( \begin{smallmatrix} n-2 \\ p \end{smallmatrix} \right) \bigl[ v^\al_{\mu \nu} \cdot k_1^p \, k_2^{n-2-p} \bigr] \, B(s+1,u+n-p-1) \, \biggr. \notag \\ 
\biggl. &\ \ + \ \Bigl[ \, (\xi^2 \,k_1) \, (u \! \not \! \xi^1)_\al \, k_3^\mu \ - \ (\xi^1 \,k_2) \, (u \! \not \! \xi^2)_\al \, k_3^\mu \ + \ (\xi^1 \, k_2) \, (u \! \not \! k_4)_\al \, \xi_2^\mu \ - \ (\xi^2 \, k_1) \, (u \! \not \! k_4)_\al \, \xi_1^\mu \Bigr. \biggr. \notag \\
\biggl. & \ \ \ \ \ \ \ \ \ \ \ \Bigl. + \ (\xi^1 \, \xi^2) \, (u \! \not \! k_1)_\al \, k_2^\mu \ - \ (\xi^1 \, \xi^2) \, (u \! \not \! k_2)_\al \, k_1^\mu \, \Bigr] \, \sum_{p=0}^{n-1} \left( \begin{smallmatrix} n-1 \\ p \end{smallmatrix} \right) \bigl[ v^\al_{\mu} \cdot k_1^p \, k_2^{n-1-p} \bigr] \, B(s,u+n-p) \biggr. \notag \\
\biggl. &\ \ + \ \Bigl[ \, \frac{1}{2} \; (u \! \not \! k_2 \! \not \! \xi^2 \! \not \! \xi^1)_\al \, k^\mu_1 \ - \ \frac{1}{2} \; (u \! \not \! k_2 \! \not \! \xi^2 \! \not \! k_1)_\al \, \xi^\mu_1 \ + \ (\xi^2 \, k_3) \, (u \! \not \! \xi^1)_\al \, k_1^\mu \ - \ (\xi^2 \, k_3)\, (u \! \not \! k_1)_\al \, \xi_1^\mu \, \Bigr. \biggr. \notag \\
\biggl. & \ \ \ \ \ \ \ \ \ \ \ \Bigl. - \ \frac{u}{2\ap} \; (u \! \not \! \xi^2)_\al \, \xi^\mu_1 \, \Bigr] \,
\sum_{p=0}^{n-1} \left( \begin{smallmatrix} n-1 \\ p \end{smallmatrix} \right) \bigl[ v^\al_{\mu} \cdot (-k_3)^p \, (-k_2)^{n-1-p} \bigr] \, B(s+n-p,u) \biggr. \notag \\
\biggl. & \ \ + \ \Bigl[ \,  \frac{1}{2} \; (u \! \not \! k_1 \! \not \! \xi^1 \! \not \! \xi^2)_\al \, k^\mu_2 \ - \ \frac{1}{2} \; (u \! \not \! k_1 \! \not \! \xi^1 \! \not \! k_2)_\al \, \xi^\mu_2 \ + \ (\xi^1 \, k_3) \, (u \! \not \! \xi^2)_\al \, k_2^\mu \ - \ (\xi^1 \, k_3)\, (u \! \not \! k_2)_\al \, \xi_2^\mu \, \Bigr. \biggr. \notag \\
\biggl. & \ \ \ \ \ \ \ \ \ \ \ \Bigl. - \ \frac{t}{2\ap} \; (u \! \not \! \xi^1)_\al \, \xi^\mu_2 \, \Bigr] \,
\sum_{p=0}^{n-1} \left( \begin{smallmatrix} n-1 \\ p \end{smallmatrix} \right) \bigl[ v^\al_{\mu} \cdot k_1^p \, (-k_3)^{n-1-p} \bigr] \, B(s+1+p,u+n-p) \, \biggr\}
\label{3,52}
\end{align} 
Expressing the Beta functions in terms of $V_t$ admits to extract the universal part of this subamplitude:
\begin{align}
{\cal A}&_{(1,1,\frac{1}{2},n+\frac{1}{2})} \ \ = \ \ \frac{\sqrt{2\ap}^{n+1} \, (n-1)! \, V_t}{\sqrt 2 \, \prod_{k=1}^{n-1}(t+k)} \notag \\
 \biggl\{ \, &\frac{1}{2\ap} \; \sum_{p=0}^{n-2} \ga(-t-1,p) \, \ga(-u-1,n-2-p) \, \xi^\mu_1 \, \xi^\nu_2 \, (u \! \not \! k_4)_\al   \, \bigl[ v^\al_{\mu \nu} \cdot (-k_1)^p \, k_2^{n-2-p} \bigr] \, \biggr. \notag \\ 
\biggl. + \ &\frac{1}{u} \, \sum_{p=0}^{n-1}  \ga(-s-1,p) \, \ga(-t-1,n-1-p)\, \bigl[ v^\al_{\mu} \cdot (-k_2)^p \, k_3^{n-1-p} \bigr]  \, \Bigl[ \, - \ \frac{u}{2\ap} \; (u \! \not \! \xi^2)_\al \, \xi^\mu_1 \Bigr. \biggr. \notag \\
\biggl. & \Bigl. \ \ \ \ \ \ + \ (\xi^2 \, k_3) \, (u \! \not \! \xi^1)_\al \, k_1^\mu \ - \ (\xi^2 \, k_3)\, (u \! \not \! k_1)_\al \, \xi_1^\mu \ + \ \frac{1}{2} \; (u \! \not \! k_2 \! \not \! \xi^2 \! \not \! \xi^1)_\al \, k^\mu_1 \ - \ \frac{1}{2} \; (u \! \not \! k_2 \! \not \! \xi^2 \! \not \! k_1)_\al \, \xi^\mu_1 \, \Bigr]  \biggr. \notag \\
\biggl. - \ &\frac{1}{t} \, \sum_{p=0}^{n-1} \ga(-s-1,p) \, \ga(-u-1,n-1-p) \,  \bigl[ v^\al_{\mu} \cdot k_1^p \, (-k_3)^{n-1-p} \bigr]  \, \Bigl[ \,- \  \frac{t}{2\ap} \; (u \! \not \! \xi^1)_\al \, \xi^\mu_2 \Bigr. \biggr. \notag \\
\biggl. & \Bigl. \ \ \ \ \ \ + \ (\xi^1 \, k_3) \, (u \! \not \! \xi^2)_\al \, k_2^\mu \ - \ (\xi^1 \, k_3)\, (u \! \not \! k_2)_\al \, \xi_2^\mu \ + \ \frac{1}{2} \; (u \! \not \! k_1 \! \not \! \xi^1 \! \not \! \xi^2)_\al \, k^\mu_2 \ - \ \frac{1}{2} \; (u \! \not \! k_1 \! \not \! \xi^1 \! \not \! k_2)_\al \, \xi^\mu_2\, \Bigr] \biggr. \notag \\
\biggl. + \ &\frac{1}{s} \,  \sum_{p=0}^{n-1} \ga(-t-1,p) \, \ga(-u-1,n-1-p)\, \bigl[ v^\al_{\mu} \cdot (-k_1)^p \, k_2^{n-1-p} \bigr]  \, \Bigl[ \, (\xi^2 \,k_1) \, (u \! \not \! \xi^1)_\al \, k_3^\mu \ - \ (\xi^1 \,k_2) \, (u \! \not \! \xi^2)_\al \, k_3^\mu   \Bigr. \biggr. \notag \\
\biggl. &\ \ \ \Bigl. + \ (\xi^1 \, k_2) \, (u \! \not \! k_4)_\al \, \xi_2^\mu \ - \ (\xi^2 \, k_1) \, (u \! \not \! k_4)_\al \, \xi_1^\mu \ + \ (\xi^1 \, \xi^2) \, (u \! \not \! k_1)_\al \, k_2^\mu \ - \ (\xi^1 \, \xi^2) \, (u \! \not \! k_2)_\al \, k_1^\mu \, \Bigr] \, \biggr\}
\label{3,53}
\end{align}
For $n=1$, this reduces to the spin $s=\frac{3}{2}$ coupling to a massless fermion and two gluons computed in \cite{lhc3}. This amplitude shares the $(-)^n$ eigenvalue under exchange of the massless bosons $\xi^1 \leftrightarrow \xi^2$ with the previous example.

\subsection{One massless boson, one higher spin boson, two massless fermions}

Here we look at another two fermi, two boson amplitude with a heavy mass level $n$ boson and both fermions massless:
\begin{align}
{\cal A}_{(\frac{1}{2},1,\frac{1}{2},n+1)} \ \ &= \ \ \langle \, cV^{(-\frac{1}{2})}(u_1,k_1) \, \int \dd z_2 \, V^{(0)} (\xi,k_2) \, cV^{(-\frac{1}{2})} (u_3 , k_3) \, c V_n^{(-1)}(\phi, k_4) \, \rangle \notag \\
&= \ \ \sqrt{2\ap}^{-n-1} \, \int \dd z_2 \ \langle \,c(z_1) \, c(z_3) \, c(z_4) \, \rangle \, \langle \, e^{-\phi(z_1) /2 }  \, e^{-\phi(z_3)/2} \, e^{-\phi(z_4) } \, \rangle \, \xi_\mu \, u_1^\al \, u_3^\be \, \phi^4_{\la_1 ... \la_n \rho} \notag \\
& \ \ \ \times \, \biggl\{ \, \langle \, i \pa X^\mu(z_2) \, i \pa X^{\la_1} \, ... \, i \pa X^{\la_n}(z_4) \prod_{j=1}^4 e^{ik_j X(z_j)} \, \rangle \, \langle \, \psi^\rho(z_4) \, S_\al(z_1) \, S_\be(z_3) \, \rangle \biggr. \notag \\
& \ \ \ \ \biggl. - \, 2 \ap \, k_\nu^2 \,  \langle \, i \pa X^{\la_1} \, ... \, i \pa X^{\la_n}(z_4) \prod_{j=1}^4 e^{ik_j X(z_j)} \, \rangle \, \langle \, \psi^\mu \, \psi^{\nu}(z_2) \, \psi^\rho(z_4) \, S_\al(z_1) \, S_\be(z_3) \, \rangle \biggr\}
\label{4,1}
\end{align}
The correlators are straightforward, and some $\ga$ matrix algebra helps to reduce the number of distinct kinematics to the following: 
\begin{align}
{\cal A}&_{(\frac{1}{2},1,\frac{1}{2},n+1)} \eq \frac{ \sqrt{2\ap}^{n+1} }{\sqrt 2} \; (-1)^{n-1} \notag \\
\biggl\{ \, &\Bigl[ \, \frac{n}{2\ap} \; (u_1 \, \ga^\mu \, u_3) \, \xi^\nu \ - \ (u_1 \, \ga^\mu \, u_3) \, k^\nu_2 \, (\xi \, k_4) \ + \ (k_2^\mu \, \xi^\la \, - \, k_2^\la \, \xi^\mu) \, (u_1 \, \ga_\la \, u_3) \,k_2^\nu \, \Bigr] \biggr. \notag \\
& \ \ \ \ \ \ \ \ \ \ \ \ \ \ \ \ \ \sum_{p=0}^{n-1} \left( \begin{smallmatrix} n-1 \\ p \end{smallmatrix} \right) \bigl[ \phi_{\mu \nu} \cdot k_1^p \, (-k_3)^{n-1-p} \bigr] \, B(s+1+p,u+n-p)  \notag \\
\biggl.  \ + \ &\Bigl[ \, (u_1 \, \ga^\mu \, u_3) \, k^\nu_1 \, (\xi \, k_3) \ + \ \frac{1}{2} \; (u_1 \, \ga^\mu \! \not \! \xi \! \not \! k_2 \, u_3) \, k_1^\nu \, \Bigr] \sum_{p=0}^{n-1} \left( \begin{smallmatrix} n-1 \\ p \end{smallmatrix} \right) \bigl[ \phi_{\mu \nu} \cdot (-k_2)^p \, (-k_3)^{n-1-p} \bigr] \, B(s+1+p,u) \biggr. \notag \\
\biggl.  \ + \ &\Bigl[ \, (u_1 \, \ga^\mu \, u_3) \, k^\nu_3 \, (\xi \, k_1) \ + \ \frac{1}{2} \; (u_1 \! \not \! k_2 \! \not \! \xi \, \ga^\mu \, u_3) \, k_3^\nu \, \Bigr] \sum_{p=0}^{n-1} \left( \begin{smallmatrix} n-1 \\ p \end{smallmatrix} \right) \bigl[ \phi_{\mu \nu} \cdot k_2^p \, k_1^{n-1-p} \bigr] \, B(s,u+1+p) \, \biggr\} 
\label{4,2}
\end{align}
The usual rearrangement of the Beta function leads to
\begin{align}
{\cal A}&_{(\frac{1}{2},1,\frac{1}{2},n+1)} \eq \frac{ \sqrt{2\ap}^{n+1} \, (n-1)! \, V_t }{\sqrt 2 \, \prod_{k=1}^{n-1}} \;  \notag \\
\biggl\{ \, &\frac{1}{t} \, \sum_{p=0}^{n-1} \ga(-s-1,p) \, \ga(-u-1,n-p-1) \, \bigl[ \phi_{\mu \nu} \cdot k_1^p \, (-k_3)^{n-1-p} \bigr] \biggr. \notag \\ 
& \ \ \ \ \ \ \ \ \ \ \ \ \ \ \ \ \ \Bigl[ \,+ \ (u_1 \, \ga^\mu \, u_3) \, k^\nu_2 \, (\xi \, k_4) \ - \ \frac{n}{2\ap} \; (u_1 \, \ga^\mu \, u_3) \, \xi^\nu \ + \ ( k_2^\la \, \xi^\mu \, - \, k_2^\mu \, \xi^\la) \, (u_1 \, \ga_\la \, u_3) \,k_2^\nu \, \Bigr]  \notag \\
\biggl. & \ + \ \frac{1}{u} \, \sum_{p=0}^{n-1} \ga(-t-1,p) \, \ga(-s-1,n-1-p) \, \bigl[ \phi_{\mu \nu} \cdot k_3^p \, (-k_2)^{n-1-p} \bigr] \biggr. \notag \\
& \ \ \ \ \ \ \ \ \ \ \ \ \ \ \ \ \ \Bigl[ \, (u_1 \, \ga^\mu \, u_3) \, k^\nu_1 \, (\xi \, k_3) \ + \ \frac{1}{2} \; (u_1 \, \ga^\mu \! \not \! \xi \! \not \! k_2 \, u_3) \, k_1^\nu \, \Bigr]  \notag \\
\biggl. & \ + \ \frac{1}{s} \, \sum_{p=0}^{n-1} \ga(-t-1,p) \, \ga(-u-1,n-1-p) \, \bigl[ \phi_{\mu \nu} \cdot k_2^{n-1-p} \, (-k_1)^{p} \bigr] \biggr. \notag \\
\biggl. & \ \ \ \ \ \ \ \ \ \ \ \ \ \ \ \ \  \Bigl[ \, (u_1 \, \ga^\mu \, u_3) \, k^\nu_3 \, (\xi \, k_1) \ + \ \frac{1}{2} \; (u_1 \! \not \! k_2 \! \not \! \xi \, \ga^\mu \, u_3) \, k_3^\nu \, \Bigr]  \, \biggr\} \ .
\label{4,3}
\end{align}
This time, one catches a sign $(-)^{n-1}$ upon exchange $u_1 \leftrightarrow u_3$, this reflects the fermionic nature of the gluinos when the $(-)^n$ from the Chan Paton trace is taken into account.

\subsection{Three massless fermions, one higher spin fermion in $D=10$}
\label{sec:4fermi}

The last four point coupling of interest involves four fermions one of which is massive with spin $s=n+\frac{1}{2}$ on the leading Regge trajectory. As we have mentioned before and explain in appendix \ref{sec:D}, four fermi amplitudes require specification of the spacetime dimension, so let us discuss the $D=10$ dimensional case here. 
\begin{align}
{\cal A}&_{(\frac{1}{2},\frac{1}{2},\frac{1}{2},n+\frac{1}{2})}^{D=10} \ \ = \ \ \langle \, cV^{(-\frac{1}{2})}(u_1,k_1) \, \int \dd z_2 \, V^{(-\frac{1}{2})} (u_2,k_2) \, cV^{(-\frac{1}{2})} (u_3 , k_3) \, c V_n^{(-\frac{1}{2})}(v,\rhob, k_4) \, \rangle \notag \\
&= \ \ \sqrt{2\ap}^{-n} \, \int \dd z_2 \ \langle \,c(z_1) \, c(z_3) \, c(z_4) \, \rangle \, \langle \, e^{-\phi(z_1) /2 } \, e^{-\phi(z_2)/2 } \, e^{-\phi(z_3)/2} \, e^{-\phi(z_4)/2} \, \rangle \, u_1^\al \, u_2^\be \, u_3^\ga \notag \\
& \ \ \ \times \, \biggl\{ \, v^\de_{\mu_1 ... \mu_n} \, \langle \, i \pa X^{\mu_1} \, ... \, i \pa X^{\mu_n}(z_4) \prod_{j=1}^4 e^{ik_j X(z_j)} \, \rangle \, \langle \,  S_\al(z_1) \, S_\be(z_2) \, S_\ga(z_3) \, S_\de(z_4) \, \rangle \biggr. \notag \\
& \ \ \ \ \ \ \biggl. + \, 2 \ap \, \rhob_{\mu_1 ... \mu_{n-1} \dde}^{\nu} \,  \langle \, i \pa X^{\mu_1} \, ... \, i \pa X^{\mu_{n-1}}(z_4) \prod_{j=1}^4 e^{ik_j X(z_j)} \, \rangle \, \langle \, S_\al(z_1) \, S_\be(z_2) \, S_\ga(z_3) \, K_\nu^{\dde}(z_4) \, \rangle \biggr\}
\label{5,1}
\end{align}
The superghost correlator $\langle e^{-\phi(z_1) /2 } e^{-\phi(z_2)/2 } e^{-\phi(z_3)/2} e^{-\phi(z_4)/2} \rangle = (z_{12} z_{13} z_{14} z_{23} z_{24} z_{34})^{-1/4}$ once again introduces fractional powers of $z_{ij}$ which are either cancelled or completed to integers by the spin field correlators from appendix \ref{appB}. By plugging all the latter into (\ref{5,1}), one arrives at
\begin{align}
&{\cal A}_{(\frac{1}{2},\frac{1}{2},\frac{1}{2},n+\frac{1}{2})}^{D=10}  \ \ = \ \ \frac{1}{2} \; \sqrt{2\ap}^n\, (-1)^{n-1} \notag \\
& \biggl\{ \, - \ \Bigl[ \, (u_1 \, \ga^\la \, u_2) \, (u_3 \, \ga_\la)_\al \,k_3^\mu \ - \ (u_1 \, \ga^\mu \, u_2)  \, (u_3 \! \not \! k_4)_\al \, \Bigr] \,
\sum_{p=0}^{n-1} \left( \begin{smallmatrix} n-1 \\ p \end{smallmatrix} \right) \, \bigl[ v_\mu^\al \cdot k_1^p \, k_2^{n-1-p} \bigr] \, B(s,u+n-p) \biggr. \notag \\
&  \ \  \biggl. - \ \Bigl[ \, (u_3 \, \ga^\la \, u_2) \, (u_1 \ga_\la)_\al \, k^\mu_1 \ - \ (u_3 \, \ga^\mu \, u_2) \, (u_1 \! \not \! k_4)_\al \, \Bigr] \, \sum_{p=0}^{n-1} \left( \begin{smallmatrix} n-1 \\ p \end{smallmatrix} \right) \, \bigl[ v_\mu^\al \cdot (-k_3)^p \, (-k_2)^{n-1-p} \bigr] \, B(s+n-p,u) \, \biggr. \notag \\
&  \ \ \biggl. + \ \Bigl[ \, (u_1 \, \ga^\la \, u_3) \, (u_2 \, \ga_\la)_\al \, k_2^\mu \ - \ (u_1 \, \ga^\mu \, u_3) \, (u_2 \! \not \! k_4)_\al \, \Bigr] \biggr. \notag \\
&  \ \ \biggl. \ \ \ \ \ \ \ \ \ \ \ \ \ \ \sum_{p=0}^{n-1} \left( \begin{smallmatrix} n-1 \\ p \end{smallmatrix} \right)  \, \bigl[ v_\mu^\al \cdot k_1^p \, (-k_3)^{n-1-p} \bigr] \, B(s+1+p,u+n-p) \, \biggr\}
\label{5,2}
\end{align}
Like in the previous examples, we reduce the beta functions to $V_t$ and find manifest cyclic symmetry in the labels $(1,2,3)$ of the massless Weyl fermions and $(-)^{n-1}$ parity under exchange of $1 \leftrightarrow 2$:
\begin{align}
{\cal A}&_{(\frac{1}{2},\frac{1}{2},\frac{1}{2},n+\frac{1}{2})}^{D=10}  \ \ = \ \ \frac{\sqrt{2\ap}^n \, (n-1)! \, V_t}{2 \, \prod_{k=1}^{n-1} (t+k)}  \notag \\
& \ \ \ \ \times \, \biggl\{ \, \frac{1}{s} \, \sum_{p=0}^{n-1} \ga(-u-1,p) \, \ga(-t-1,n-1-p) \biggr. \notag \\
& \ \ \ \ \ \ \ \ \ \ \ \ \ \ \ \ \ \ \ \ \ \ \ \ \biggl. \Bigl[ \, (u_1 \, \ga^\mu \, u_2)  \, (u_3 \! \not \! k_4)_\al \ - \ (u_1 \, \ga^\la \, u_2) \, (u_3 \, \ga_\la)_\al \, k^\mu_3 \, \Bigr] \, \bigl[ v^\al_\mu \cdot k_2^{p} \, (-k_1)^{n-p-1} \bigr]  \biggr. \notag \\
& \ \ \ \ \ \ \ \ \ \biggl. + \ \frac{1}{t} \, \sum_{p=0}^{n-1} \ga(-s-1,p) \, \ga(-u-1,n-1-p) \biggr. \notag \\
& \ \ \ \ \ \ \ \ \ \ \ \ \ \ \ \ \ \ \ \ \ \ \ \ \biggl. \Bigl[ \, (u_1 \, \ga^\mu \, u_3)  \, (u_2 \! \not \! k_4)_\al \ - \ (u_1 \, \ga^\la u_3) \, (u_2 \, \ga_\la)_\al \, k^\mu_2 \, \Bigr] \, \bigl[ v^\al_\mu \cdot k_1^{p} \, (-k_3)^{n-p-1} \bigr]  \biggr. \notag \\
& \ \ \ \ \ \ \ \ \ \biggl. + \ \frac{1}{u} \, \sum_{p=0}^{n-1} \ga(-t-1,p) \, \ga(-s-1,n-1-p) \biggr. \notag \\
& \ \ \ \ \ \ \ \ \ \ \ \ \ \ \ \ \ \ \ \ \ \ \ \ \biggl. \Bigl[ \, (u_3 \, \ga^\mu \, u_2)  \, (u_1 \! \not \! k_4)_\al \ - \ (u_3 \, \ga^\la \, u_2) \, (u_1 \ga_\la)_\al \, k^\mu_1 \, \Bigr] \, \bigl[ v^\al_\mu \cdot k_3^{p} \, (-k_2)^{n-p-1} \bigr] \, \biggr\} 
\label{5,3}
\end{align}
Let us display the simplest case $n = 1$ explicitly because \cite{lhc3} only gives the $D=4$ result with chiral fermions:
\begin{align}
{\cal A}&_{(\frac{1}{2},\frac{1}{2},\frac{1}{2},\frac{3}{2})}^{D=10} \eq \sqrt{ \frac{ \ap}{2} } \; V_t \, \biggl\{ \, \frac{1}{s} \; (u_1 \, \ga^\mu \, u_2) \, \Bigl[ \, (u_3 \! \not \! k_4 \, v_\mu) \ - \ (u_3 \, \ga_\mu \, v^\nu) \, k_\nu^3 \, \Bigr] \biggr. \notag \\
& \ \ \biggl. + \ \frac{1}{u} \; (u_2 \, \ga^\mu \, u_3) \, \Bigl[ \, (u_1 \! \not \! k_4 \, v_\mu) \ - \ (u_1 \, \ga_\mu \, v^\nu) \, k_\nu^1\, \Bigr] \ + \ \frac{1}{t} \; (u_3 \, \ga^\mu \, u_1) \, \Bigl[ \, (u_2 \! \not \! k_4 \, v_\mu) \ - \ (u_2 \, \ga_\mu \, v^\nu) \, k_\nu^2 \, \Bigr] \, \biggr\}
\label{5,4}
\end{align}

%

%

\section{Conclusion and further directions}

In this article, we have computed three-point vertices of leading Regge trajectory states with arbitrary (integer of half-integer) spin. Our finding (\ref{2,7}) for the three boson coupling can be regarded as the superstring generalization of equation (4.1) in \cite{bos} for bosonic string theory, and (\ref{f2}) gives the fermionic completion for which \cite{bos} provides an educated guess in the massless $\ap \rightarrow \infty$ regime. 

\medskip
We got a glimpse of four-point interactions by coupling higher spins to three massless particles. The first mass level cases $n=1$ were already known \cite{lhc3} and discussed in a phenomenological context. In this work, we have generalized to production amplitudes for higher mass Regge excitations at level $n \geq 2$. Our four-point results are expressed in terms of the Euler Beta functions $B(s+p,u+q)$ of Mandelstam invariants $s$ and $u$ emerging from the worldsheet integrals with positive integer shifts $p,q \in \NN_0$. Furthermore, a presentation in terms of the stringy formfactor $V_t$ (with trivial field theory limit $V_t \rightarrow 1$ as $\ap \rightarrow 0$) is suitable to make the amplitudes' symmetry more manifest,  in particular cyclic invariance in case of three identical massless states.

\medskip 
Both our three- and four-point results are valid in any dimension $D \leq 10$ except for the four fermion amplitude in subsection \ref{sec:4fermi}.

\medskip
There are many interesting directions into which the present analysis should be extended. Along the lines of \cite{bos}, it would be important to find generating functions for these couplings \cite{future}. Moreover, it is left for future work to extract the massless limit and to see recovery of gauge symmetry in the low tension limit. From explicit results for cubic and higher order interactions, off-shell currents that lie behind string interactions wait to be identified. Off shell vertices in bosonic string theory are constructed in the updated version of \cite{bos} and in \cite{Fotopoulos:2010ay}, also see \cite{Manvelyan:2010je} for generalizations. Finally, our results should be analyzed in view of jet signals in low string scale scenarios \cite{lhc1,lhc2,lhc3}.


%
\vskip1cm
\goodbreak
\centerline{\noindent{\bf Acknowledgments} }\vskip 2mm

I wish to thank Alexander Dobrinevski, Robert Richter, Stephan Stieberger and Massimo Taronna for helpful discussion and for their valuable suggestions to improve earlier versions of this paper. I am grateful to Augusto Sagnotti for encouragements and continuous email correspondence. Karapet Mkrtchyan and Eugene Skvortsov pointed out important references which are now included in from the second arXiv version on. Also, I would not have wanted to miss the inspiring coffee breaks with Alois Kabelschacht.

\section*{Appendix}
\appendix

\section{Fermions in $D<10$ dimensions}
\label{sec:D}

The purpose of this appendix is to sketch the implementation of compactification on the level of fermionic vertex operators and to give the argument why any considered scattering amplitude with two fermions does not depend on the number $D$ of uncompactified dimensions. It closes the gap which was left in subsection \ref{sec:dim}.

\medskip
The impact of compactification on the fermion vertex is a bit more involved than for its bosonic cousin: Of course, vector indices $\mu_i$ of the fermionic wavefunctions $v_{\mu_1...\mu_n}^\al$ and $\rhob_{\mu_1... \mu_n \dbe}$ simply have to be restricted to their first $D$ values $\mu_i \equiv m_i$ in order to preserve spin $s=n+\frac{1}{2}$ under $SO(1,D-1)$. But the $D$ dimensional remanent of Weyl spinors $(v^\al,\rhob_{\dbe})$ requires more care. In Cartan Weyl basis, the values of Weyl spinor indices $_\al$ ($^{\dbe}$) can be thought of as weight vectors $\left( \pm \frac{1}{2}, \pm \frac{1}{2}, \pm \frac{1}{2}, \pm \frac{1}{2}, \pm \frac{1}{2} \right)$ with an even (odd) number of minus signs, each entry being associated with a pair of the initially ten spacetime dimensions. To take the decoupling of $2n=10-D$ compactified dimensions from the uncompactified $SO(1,D-1)$ into account, one needs to freeze the last $n$ entries of the $_\al,^{\dbe} \leftrightarrow \left( \pm \frac{1}{2}, \pm \frac{1}{2}, \pm \frac{1}{2}, \pm \frac{1}{2}, \pm \frac{1}{2} \right)$ vector.

\medskip
In the RNS CFT, the spin field $S_\al$ is factorized as $S_a \otimes s_{\te{int}}^i$ 
into an internal part $s^{i}_{\te{int}}$ and a spinor $S_a$ 
in $D$ dimensions 
whose indices can be represented by $D/2$ component weight vectors $_a \leftrightarrow \left( \pm \frac{1}{2}, ..., \pm \frac{1}{2} \right)$
. The conformal dimension $\frac{5}{8}$ of $S_\al$ is split into $h_1=\frac{D}{16}$ for the spacetime part $S_a$ 
and $h_2=\frac{10-D}{16}$ for the internal spin field, analogous statements hold for the composite operator $K_\mu^{\dbe} \equiv \psi_\mu  \! \not \! \! \psi^{\al \dbe} S_\al$ with $SO(1,D-1)$ index $\mu \equiv m$. Two point functions of the $s_{\te{int}}^{i}$ fields are completely determined by their conformal weight,
\beq
\langle \, s_{\te{int}}^i(z) \, s_{\te{int}}^j(w) \, \rangle \eq \frac{c^{ij}}{(z-w)^{(10-D)/8}} \co c^{ij} \in \{ 0,1 \} \, \equiv \, \te{generalized charge conjugation matrix} \ ,
\label{comp1}
\eeq
and this is the only signature of the internal dimensions' CFT in two fermion amplitudes with exclusively leading Regge trajectory states since internal components of the $\psi^\mu$ which might interact with $s_{\te{int}}^{i}$ are forbidden by the maximum spin requirement. Correlations of spacetime fields $\psi^m$ and $S_a$, 
on the other hand, are made up of the $D$ dimensional gamma matrices $\ga^m_{ab}$. 

\medskip
To give an easy and still illuminating example of how two fermion amplitude might become independent on $D$, let us consider the three point function of the $SO(1,D-1)$ spin fields
\beq
\langle \, \psi^m(z_1) \, S_a(z_2) \, S_{b}(z_3) \, \rangle \eq \frac{ \ga^m_{a b}}{\sqrt{2} \, (z_{12} \, z_{13})^{1/2} \, z_{23}^{D/8 - 1/2}} \ .
\label{comp2}
\eeq
The complementary factor of $z_{23}^{D/8}$ within (\ref{comp1}) makes sure that the overall $D$ dependence drops out of ten dimensional spin field correlation
\beq
\langle \, \psi^m(z_1) \, S_\al(z_2) \, S_\be(z_3) \, \rangle \, \Bigl. \Bigr|_{D<10} \eq \langle \, \psi^m(z_1) \, S_a\, s_{\te{int}}^i (z_2) \, S_{ b}\, s_{\te{int}}^j(z_3) \, \rangle \eq \frac{ \ga^m_{a  b} \, c^{ij}}{\sqrt{2} \, (z_{12} \, z_{13})^{1/2} \, z_{23}^{3/4}} \ .
\label{comp3}
\eeq
The same mechanism leads to $D$ independent $z_{ij}$ powers in higher point correlators with $\psi^\mu,S_\al,K_\mu^{\dbe}$ fields. The only potential $D$ dependence might enter from the spacetime kinematics within the $K_\mu^{\dbe}$ correlation functions of appendix \ref{appB}. But the spin $3/2$ projection $\bar \rho_{\dbe}^\mu K_\mu^{\dbe}$ (with $\ga$ traceless wavefunction $\bar \rho$ as it appears on the leading trajectory) produces a $(D-2)$ factor in all its worldsheet interactions, see the OPEs (\ref{OPE2}), (\ref{OPE3}) and (\ref{OPE5}). The $(D-2)$ factors in correlation functions with $\bar \rho_{\dbe}^\mu K_\mu^{\dbe}$ can then be removed by means of the onshell constraint $(D-2) \rhob_{\dbe} = -v^\al \!\not \! k_{\al \dbe}$. This mechanism therefore holds for amplitudes with two fermions and any number of bosons.

\medskip
One comment on the $SO(1,D-1)$ chirality of $S_a$ spin fields and the associated wave function $v^a$ is in order: Because of the different chirality structure in the charge conjugation matrix for dimensions $(2 \ \te{mod} \ 4)$ and $(4 \ \te{mod} \ 4)$, the correlator (\ref{comp2}) vanishes for alike $(S_a,S_b)$ chiralities if $D=4 \ \te{mod} \ 4$ and for opposite chiralities in $D=2 \ \te{mod} \ 4$ dimensional compactifications. The relative $D$ dimensional chirality of the spinor wave functions in compactified two fermion amplitudes ${\cal A}_{(s_1,n_2+\frac{1}{2},n_3+\frac{1}{2})}$, ${\cal A}_{(1,1,\frac{1}{2},n+\frac{1}{2})}$ and ${\cal A}_{(\frac{1}{2},1,\frac{1}{2},n+1)}$ must therefore be alike if $D=2 \ \te{mod} \ 4$ and opposite if $D=4 \ \te{mod} \ 4$. In order to keep the notation outside this appendix simple, we will use $^\mu, _\al, ^{\dbe}$ indices rather than $^m,_a,^{\dot b}$ for the correlators in appendix \ref{appB} although they are given for general dimensions $D$.

\medskip
To give a few more details about the internal spin fields: In a maximally supersymmetric compactification, they can be represented by simple exponentials of $m = \frac{10-D}{2}$ free bosons $s_{\te{int}}^{(\pm,...,\pm)} = \ee^{\frac{i}{2}(\pm H_1,...,\pm H_m)}$, multiplied by an $m$ component weight vector of $SO(10-D)$. If some supersymmetries are broken, certain $\pm$ choices and therefore gaugino species are projected out. To describe chiral matter in orbifolds or intersecting brane models, on the other hand, the $s_{\te{int}}^i$ can be attributed to the twisted CFT sector and carry details of the compactification geometry such as brane intersection angles. In this case, an appropriate generalization $c^{ij}$ of an internal charge conjugation matrix enters the two point functions (\ref{comp1}). If this is nonzero, then any two fermion amplitude with otherwise bosonic highest spin states cannot distinguish between adjoint fermions from the untwisted sector, say gauginos, and chiral matter (e.g. quarks and leptons) with a boundary changing vertex operator.

\medskip
Four fermion amplitudes require four point functions $\langle s_{\te{int}}^i(z_1) s_{\te{int}}^j(z_2) s_{\te{int}}^k(z_3) s_{\te{int}}^l(z_4) \rangle$ as a CFT input which cannot be discussed in a model independent fashion. They can have highly non-trivial dependence on cross ratios of the worldsheet positions, parametrized by data of the internal geometry. Four quark scattering in \cite{lhc1,lhc2,lhc3} provides examples of how boundary changing operators for brane intersections give rise to amplitudes depending on intersection angles. 

\medskip
But even if we stick to the simplest realization of fermions, for instance with only the two internal spin fields $s_{\te{int}}^{\pm} = \ee^{\pm \frac{i}{2}(H_1,..., H_m)}$ available for the unique gaugino species in ${\cal N}=1$ supersymmetry, one cannot write down $D$ independent expressions for quartic fermi couplings for representation theoretic reasons: The structure of Fierz identities and therefore the appropriate bases of Lorentz tensors vary a lot with $D$. In $D=4$ and $D=6$, there exist three linearly independent tensors with four free spinor indices whereas $D=8$ and $D=10$ admit five independent such tensors.\footnote{Explicitly, these are:
\[ \begin{array}{cl}
C_{\al \be} \, C_{\ga \de} \ , \ \ \ C_{\al  \de} \, C_{\ga \be} \ , \ \ \ C_{\al \be} \, C^{\dga \dde} &: \ D=4 \\
\tfrac{1}{2} \, (\ga^\mu \, C)_{\al \be} \, (\ga_\mu \, C)_{\ga \de} \ = \ \vep_{\al \be \de \ga} \ , \ \ \ C_\al{}^{\dga} \, C_\be{}^{\dde} \ , \ \ \ C_\al{}^{\dde} \, C_\be{}^{\dga} &: \ D=6 \\
C_{\al \be} \, C_{\ga \de} \ , \ \ \ C_{\al  \de} \, C_{\ga \be} \ , \ \ \ C_{\al \ga} \, C_{\be \de} \ , \ \ \ (\ga^\mu \, C)_\al {}^{\dga} \, (\ga_\mu \, C)_\be {}^{\dde} \ , \ \ \ (\ga^\mu \, C)_\al {}^{\dde} \, (\ga_\mu \, C)_\be {}^{\dga} &: \ D=8 \\
(\ga^\mu \, C)_{\al \be} \, (\ga_\mu \, C)_{\ga \de} \ , \ \ \ (\ga^\mu \, C)_{\al \de} \, (\ga_\mu \, C)_{\ga \be} \ , \ \ \ (\ga^\mu \, C)_{\al \be} \, (\gab_\mu \, C)^{\dga \dde} \ , \ \ \ C_\al{}^{\dga} \, C_\be{}^{\dde} \ , \ \ \ C_\al{}^{\dde} \, C_\be{}^{\dga} &: \ D=10
\end{array} \]
Any higher order $\ga$ matrix contraction can be reduced to the shown tensors by means of Fierz identities.}

\section{CFT correlation functions for bosonic states}
\label{appA}

The scattering amplitudes of higher spin states on the leading Regge trajectory involve manageable correlation functions of the NS fermion $\psi^\mu$. Specifically, we need
\begin{align}
\langle \, \psi^{\tau}(z_2) \, \psi^{\si}(z_3) \, \rangle \ \ &= \ \ \frac{\eta^{\tau \si}}{z_{23}} \notag \\
\langle \, \psi^{\al}(z_1) \, \psi^{\rho}(z_1) \, \psi^{\tau}(z_2) \, \psi^{\si}(z_3) \, \rangle \ \ &= \ \ \frac{ \eta^{\rho \tau} \, \eta^{\al \si} \, - \, \eta^{\al \tau} \, \eta^{\rho \si} }{z_{12} \, z_{13} } \label{A1} \\
\langle \, \pa \psi^{\mu_{n_1}}(z_1) \, \psi^{\rho}(z_1) \, \psi^{\tau}(z_2) \, \psi^{\si}(z_3) \, \rangle \ \ &= \ \ \frac{ \eta^{\mu_{n_1} \tau} \, \eta^{\rho \si}}{z_{12}^2 \, z_{13}} \ - \ \frac{ \eta^{\mu_{n_1} \si} \, \eta^{\rho \tau}}{z_{12} \, z_{13}^2} \notag
\end{align}
for the three point vertex as well as the following additional one for four point amplitudes:
\begin{align}
\langle \, \psi^\mu \, &\psi^\nu(z_1) \, \psi^\ka(z_2) \, \psi^\la \, \psi^\rho(z_3) \, \psi^\xi(z_4) \, \rangle \eq \frac{ \eta^{\ka \xi} \, (\eta^{\nu \la} \, \eta^{\mu \rho} \, - \, \eta^{\mu \la} \, \eta^{\nu \rho}) }{ z_{13}^2 \, z_{24} } \notag \\
& + \ \frac{ \eta^{\mu \la} \, \eta^{\nu \ka} \, \eta^{\rho \xi} \, - \, \eta^{\nu \la} \, \eta^{\mu \ka} \, \eta^{\rho \xi} \, - \, \eta^{\mu \rho} \, \eta^{\nu \ka} \, \eta^{\la \xi} \, + \, \eta^{\nu \rho} \, \eta^{\mu \ka} \, \eta^{\la \xi} }{z_{12} \, z_{13} \, z_{34} } \notag \\
& + \ \frac{ \eta^{\mu \la} \, \eta^{\nu \xi} \, \eta^{\rho \ka} \, - \, \eta^{\nu \la} \, \eta^{\mu \xi} \, \eta^{\rho \ka} \, - \, \eta^{\mu \rho} \, \eta^{\nu \xi} \, \eta^{\la \ka} \, + \, \eta^{\nu \rho} \, \eta^{\mu \xi} \, \eta^{\la \ka} }{z_{13} \, z_{14} \, z_{23}}
\label{A2}
\end{align}
Correlations of $i\pa X$ and $e^{ikX}$ fields represent the tougher part of the CFT challenge when scattering spacetime bosons. Luckily, lots of simplifications arise because the $i \pa X^\mu$ are always contracted with the totally symmetric, transverse and traceless wave functions $\phi$, this is why we only give a closed formula for such a contraction. Let $n_i$ denote the mass levels in the sense that $k_i^2 = -\frac{n_i}{\ap}$ and $s_i$ the number of $i\pa X(z_i)$'s (which does not necessarily coincide with the spin $n_i+1$ of the state for the purpose of this appendix). Of course, the tensors $\phi_{\mu_1 ... \mu_s}$ of interest might have further free indices which will be omitted in the following result:
\begin{align}
\phi^1_{\mu_1 ... \mu_{s_1}} &\, \phi^2_{\nu_1 ... \nu_{s_2}} \, \phi^3_{\la_1 ... \la_{s_3}} \, \langle \, \prod_{p_1=1}^{s_1} \, i \pa X^{\mu_{p_1}}(z_1) \, \prod_{p_2=1}^{s_2} \, i \pa X^{\nu_{p_2}}(z_2) \, \prod_{p_3=1}^{s_3} \, i \pa X^{\la_{p_3}}(z_3) \,  \prod_{j=1}^3 e^{ik_j X(z_j)} \, \rangle \notag \\
&= \ \ \left( \frac{ z_{12} \, z_{13} }{ z_{23} } \right)^{n_1-s_1} \, \left( \frac{ z_{12} \, z_{23} }{ z_{13} } \right)^{n_2-s_2} \, \left( \frac{ z_{13} \, z_{23} }{ z_{12} } \right)^{n_3-s_3} \, s_1! \, s_2! \, s_3! \sum_{i,j,k \in {\cal I}} (2\ap)^{s_1+s_2+s_3-i-j-k} \notag \\
& \ \ \ \ \ \ \ \ \times \ \frac{ ( \phi^1 \cdot k_2^{s_1-i-j}) \, (\phi^2 \cdot k_3^{s_2-i-k}) \, (\phi^3 \cdot k_1^{s_3-j-k} ) \, \de_{12}^i \, \de_{13}^j \, \de_{23}^k }{i! \, j! \, k! \, (s_1 - i - j)! \, (s_2 - i - k)! \, (s_3 - j-k)!}
\label{A3}
\end{align}
See subsection \ref{sec:3pt} for the shorthands such as $( \phi^1 \cdot k_2^{s_1-i-j}) \, (\phi^2 \cdot k_3^{s_2-i-k}) \, \de_{12}^i$. The summation range ${\cal I}$ for the number $i,j,k$ of contractions among the $\phi$'s is defined as
\beq
{\cal I} \ \ := \ \ \Bigl\{  \, i,j,k \in \NN_0 \, : \ s_1 - i - j \geq 0 \ , \ \ \ s_2 - i-k \geq 0 \ , \ \ \ s_3 - j - k \geq 0 \, \Bigr\} \ .
\label{A4}
\eeq
The four particle generalization is given as follows:
\begin{align}
&\phi^1_{\mu_i} \, \phi^2_{\nu_i} \, \phi^3_{\la_i} \, \phi^4_{\rho_i} \, \langle \, \prod_{p_1=1}^{s_1} \, i \pa X^{\mu_{p_1}}(z_1) \, \prod_{p_2=1}^{s_2} \, i \pa X^{\nu_{p_2}}(z_2) \, \prod_{p_3=1}^{s_3} \, i \pa X^{\la_{p_3}}(z_3) \, \prod_{p_4=1}^{s_4} \, i \pa X^{\rho_{p_4}}(z_4) \,  \prod_{j=1}^4 e^{ik_j X(z_j)} \, \rangle \notag \\
&= \ \ | z_{12} |^{s+n_1+n_2} \, | z_{13} |^{t+n_1+n_3} \, | z_{14} |^{u+n_1+n_4} \, | z_{23} |^{u+n_2+n_3} \, | z_{24} |^{t+n_2+n_4} \, | z_{34} |^{s+n_3+n_4}  \, s_1! \, s_2! \, s_3! \, s_4! \notag \\
& \  \times \! \! \! \! \! \sum_{i,j,k,l,m,n \in {\cal J}} \! \! \! \! \! \frac{(2\ap)^{s_1+s_2+s_3+s_4-i-j-k-l-m-n} \ (\de_{12} / z_{12}^{2})^i \ (\de_{13} / z_{13}^{2})^j \ (\de_{23} / z_{23}^{2})^k \ (\de_{14} / z_{14}^{2})^l \ (\de_{24} / z_{24}^{2} )^m \ (\de_{34} / z_{34}^{2})^n}{i! \, j! \, k! \, l! \, m! \, n! \, (s_1 - i - j-l)! \, (s_2 - i - k-m)! \, (s_3 - j-k-n)! \, (s_4-l-m-n)!} \notag \\
& \  \times \ \left[ \, \phi^1 \cdot  \left( \frac{k_2}{z_{12}} \, + \, \frac{k_3}{z_{13}} \, + \, \frac{k_4}{z_{14}} \right)^{s_1-i-j-l} \, \right] \ \left[ \, \phi^2 \cdot  \left( \frac{k_1}{z_{21}} \, + \, \frac{k_3}{z_{23}} \, + \, \frac{k_4}{z_{24}} \right)^{s_2-i-k-m} \, \right] \notag \\
& \  \times \ \left[ \, \phi^3 \cdot  \left( \frac{k_1}{z_{31}} \, + \, \frac{k_2}{z_{32}} \, + \, \frac{k_4}{z_{34}} \right)^{s_3-j-k-n} \, \right] \ \left[ \, \phi^4 \cdot  \left( \frac{k_1}{z_{41}} \, + \, \frac{k_2}{z_{42}} \, + \, \frac{k_3}{z_{43}} \right)^{s_4-l-m-n} \, \right]
\label{A5}
\end{align}
In this case, summation variables $i,j,k,l,m,n$ are delimited as follows:
\beq
{\cal J} \ \, := \, \ \Bigl\{  \, i,j,k,l,m,n \in \NN_0 \, : \ s_1 - i - j -l \geq 0 \ , \ \ \ s_2 - i-k-m \geq 0 \ , \ \ \ s_3 - j - k - n \geq 0 \ , \ \ \ s_4 -l-m-n \geq 0 \, \Bigr\} 
\label{A6}
\eeq
The four point amplitudes given in section \ref{sec:4pt1} with one higher spin state of mass level $n$ at $z_4$ and massless states at $z_{1,2,3}$ require
\begin{align}
\phi_{\mu_1 ... \mu_{s}} \, \langle \, i\pa X^{\mu_1} \, ... \, i \pa X^{\mu_{s}}(z_4) \,  \prod_{j=1}^4 e^{ik_j X(z_j)} \, &\rangle \eq | z_{12} |^{s} \, | z_{13} |^{t} \, | z_{23} |^{u} \, | z_{14} |^{u+n} \,  | z_{24} |^{t+n} \, | z_{34} |^{s+n} \notag \\
&\ \ \ \ \ \ \ \ \ \ \ \ \ \ \ \ \ \times \ \left[ \, \phi^4 \cdot  \left( \frac{k_1}{z_{41}} \, + \, \frac{k_2}{z_{42}} \, + \, \frac{k_3}{z_{43}} \right)^{s} \, \right]\ .
\label{A10}
\end{align}

\section{CFT correlation functions for fermionic states}
\label{appB}

In this appendix, we list a choice of correlation functions containing spin fields $S_\al$ and the composite operators $K_\mu^{\dal} = \psi_\mu \psi^\nu \gab_\nu^{\dal \be} S_\be$ of conformal dimension $h=1 + \frac{D}{16}$ which appear in the massive fermion vertex. Correlations of two spinorial fields will be given for general spacetime dimension $D$ (although dimensions $D=2 \, \te{mod} \, 4$ strictly speaking require a different relative chirality of the spinors than $D=4 \, \te{mod} \, 4$), see appendix \ref{sec:D}. In presence of four or more spin fields, however, the spacetime dimensions will be fixed to $D=10$ since the structure of Fierz identities and therefore the appropriate bases of Lorentz tensors vary a lot with $D$.

\medskip
The interaction of spin field with NS fermions was studied in various dimensions \cite{spin1,spin2,spin3} -- both at tree level and on higher genus. The relevant correlators for amplitudes in this work are
\begin{align}
\langle \, \psi^\mu(z_1) \, S_\al(z_2) \, S_\be(z_3) \, \rangle \ \ &= \ \ \frac{(\ga^\mu \, C)_{\al \be}}{\sqrt{2} \, (z_{12} \, z_{13})^{1/2} \, z_{23}^{D/8- 1/2}} \label{S1} \\
\langle \, \psi^\mu(z_1) \, \psi^\nu \, \psi^\la(z_2) \, S_\al(z_3) \, S_\be(z_4) \, \rangle \ \ &= \ \ \frac{1}{\sqrt{2} \, (z_{13} \, z_{14})^{1/2} \, z_{23} \, z_{24} \, z_{34}^{D/8- 1/2}} \;\biggl\{ \, \frac{z_{34}}{2} \; (\ga^{\mu} \, \bar \ga^{\nu \la} \, C)_{\al \be} \biggr. \notag \\
& \  \ \ \ \ \ \ \ \ \ \  \ \ \biggl. + \ \frac{\eta^{\mu \nu} \, (\ga^\la \, C)_{\al \be} \ - \ \eta^{\mu \la} \, (\ga^\nu \, C)_{\al \be}}{z_{12}} \; z_{13} \, z_{24} \, \biggr\}
\label{S2} \\
\langle \, S_\al(z_1) \, S_\be(z_2) \, S_\ga(z_3) \, S_\de(z_4) \, \rangle \, \Bigl. \Bigr|_{D=10} \ \ &= \ \ \frac{(z_{12} \, z_{14} \, z_{23} \, z_{34})^{1/4}}{2 \, (z_{13} \, z_{24})^{3/4}} \; \biggl\{ \, \frac{ (\ga^\mu \, C)_{\al \be} \, (\ga_\mu \, C)_{\ga \de}}{z_{12} \, z_{34}} \ - \ \frac{ (\ga^\mu \, C)_{\al \de} \, (\ga_\mu \, C)_{\ga \be}}{z_{14} \, z_{23}} \, \biggr\} \notag \\
= \ \ \frac{1}{6 \, (z_{12} \, z_{13} \, z_{14} \, z_{23} \, z_{24} \, z_{34})^{3/4}} \; &\biggl\{ \, (\ga^\mu \, C)_{\al \be} \, (\ga_\mu \, C)_{\ga \de} \, (z_{13} \, z_{24} \ + \ z_{14} \, z_{23}) \biggr. \notag \\
\biggl. + \ (\ga^\mu \, C)_{\al \ga} \, (\ga_\mu \, C)_{\be \de} \, &(z_{12} \, z_{34} \ - \ z_{14} \, z_{23}) \ - \ (\ga^\mu \, C)_{\be \ga} \, (\ga_\mu \, C)_{\al \de} \, (z_{12} \, z_{34} \ + \ z_{13} \, z_{24}) \, \biggr\}
\label{S3}
\end{align}
Correlation functions with $K_\mu^{\dal}$ are very involved in general. But in the context of leading regge trajectory fermions, they are contracted by a $\ga$ traceless wavefunction $\bar \rho^\mu_{\dal}$ such that only their spin $3/2$ irreducible contributes. This spin $3/2$ projection simplifies the correlators enormously, so we will always give their $\bar \rho$ contractions in the following.

\medskip
The spin $3/2$ component of $K_\mu^{\dal}$ is governed by the following OPEs:
\begin{align}
\bar \rho^\mu_{\dbe} \, S_\al(z) \, K_\mu^{\dbe}(w) \ \ &\sim \ \ \frac{(D-2) \, C_\al {}^{\dbe} }{\sqrt{2} \, (z-w)^{1/2 + D/8}} \; \bar \rho^\mu_{\dbe} \, \psi_\mu(w) \label{OPE2} \\
\bar \rho^\nu_{\dal} \, \psi^\mu(z) \, K_\nu^{\dal}(w) \ \ &\sim \ \ \frac{ (D-2)  }{\sqrt{2} \, (z-w)^{3/2}} \; \bar \rho^\mu_{\dal} \, S^{\dal}(w) \label{OPE3} \\
\bar \rho^{\mu}_{\dal} \, \bar \rho^{\nu}_{\dbe} \, K_\mu^{\dal}(z) \, K_\nu^{\dbe}(w) \ \ &\sim \ \ \frac{(D-2)^2}{2 \sqrt{2} \, (z-w)^{3/2 + D/8}} \; \bar \rho^\mu_{\dal } \, (\gab_\la \, C)^{\dal \dbe} \, \bar \rho_{\mu \dbe} \, \psi^\la(w) \label{OPE5}
\end{align}
Given these prerequisites, one can verify the correct singular behaviour of 
the three point functions
\begin{align}
\bar \rho^\nu_{\dbe} \, \langle \, \psi^{\mu}(z_1) \, S_{\al}(z_2) \, K_\nu^{\dbe}(z_3) \, \rangle \ \ &= \ \ \frac{(D-2) \, z_{12}^{1/2}}{\sqrt{2} \, z_{13}^{3/2} \, z_{23}^{D/8 + 1/2}}  \; \bar \rho^{\mu }_{\dbe} \, C_\al {}^{\dbe}  \label{K3} \\
\bar \rho^\nu_{2 \dal} \, \bar \rho^\la_{3\dbe} \, \langle \, \psi^{\mu}(z_1) \, K_\nu^{\dal}(z_2) \, K_\la^{\dbe}(z_3) \, \rangle \ \ &= \ \ \frac{(D-2)^2}{2\sqrt{2} \, (z_{12} \, z_{13})^{1/2} \, z_{23}^{D/8 + 3/2}}  \; \bar \rho^{\nu}_{2 \dal}   \, (\gab^\mu \, C)^{\dal \dbe} \, \bar \rho_{3 \nu \dbe} \label{K4}
\end{align}
as well as the (no longer $D$ universal) four point function
\begin{align}
\bar \rho^\mu_{\dde} \, \langle \, S_\al(z_1) \, &S_\be(z_2) \, S_\ga(z_3) \, K_\mu^{\dde}(z_4) \, \rangle \, \Bigl. \Bigr|_{D=10} \ \ = \ \ \frac{4 \, (z_{12}\, z_{13} \, z_{14} \, z_{23} \, z_{24} \, z_{34})^{1/4} \, \bar \rho^\mu_{\dde}}{z_{14} \, z_{24} \, z_{34} }  \notag \\
\times \, \biggl\{ \, &\frac{ (\ga^\mu \, C)_{\al \be} \,C_\ga{}^{\dde} }{z_{12} \, z_{34} } \ - \ \frac{ (\ga^\mu \, C)_{\al \ga} \,C_\be{}^{\dde} }{z_{13} \, z_{24} }  \ + \ \frac{ (\ga^\mu \, C)_{ \be \ga } \,C_\al{}^{\dde} }{z_{14} \, z_{23} } \, \biggr\}  \ .
 \label{K5}
\end{align}
Also, we need the following five point correlator for the scattering of two gluons with a massless and a massive fermion:
\begin{align}
\bar \rho^\tau_{\dbe} \, \langle \, \psi^\mu(z_1) \, &\psi^{\nu}(z_2) \, \psi^{\la}(z_2) \, S_{\al}(z_3) \, K^{\dbe}_{\tau}(z_4) \, \rangle \ \ = \ \ \frac{(D-2) \, \bar \rho^\tau_{\dbe}}{ \sqrt{2} \, (z_{13} \, z_{14})^{1/2} \, z_{23} \, z_{24} \, z_{34}^{3/4}} \notag \\
\biggl\{ \, &\frac{ (\eta^{\mu \nu} \, \de^{\la}_{\tau} \ - \ \eta^{\mu \la} \, \de^{\nu}_{ \tau }) \, C_\al {}^{\dbe} }{ z_{12} \,  z_{34}} \; z_{23} \, z_{13} \ + \ \frac{ \de^{\mu }_{\tau} \, (\ga^{\nu \la} \, C)_{\al} {}^{\dbe} }{2 \, z_{14} } \; z_{13} \biggr. \notag \\
\biggl. & \ \ \ + \ \frac{ \de^{\la}_{ \tau} \, (\ga^{\mu} \, \gab^{\nu} \, C)_\al{}^{\dbe} \ - \ \de^{\nu}_{ \tau } \, (\ga^{\mu} \, \gab^{ \la} \, C)_\al{}^{\dbe} }{2 \, z_{24}} \; z_{23} \, \biggr\} 
\end{align}
In all scattering amplitudes given in this paper, the $\bar \rho$ wavefunctions were eliminated by means of the massive Dirac equation $(D-2) \bar \rho_{\dal} = - v^{\al} \not \! k_{\al \dal}$ (suppressing any free vector index).

\section{Dirac spinor conventions}
\label{appC}

This appendix gives the detailed dictionary between Dirac spinors used in subsection \ref{sec:Dirac} and their irreducible Weyl components. Left handed spinor indices $_\al$ and right handed ones $^{\dal}$ will collectively be denoted as $_A \in \{ _\al , ^{\dal} \}$ or $^A \in \{ ^\al , _{\dal} \}$, and the symmetrized vector indices of our spin $n+\frac{1}{2}$ wave functions $v,\rhob, \bar w$ will be suppressed in this appendix.

\medskip
In order to align the left handed fermion wavefunctions $v^\al$ into a Dirac spinor $\Psi^A$ subject to the Dirac equation at mass $m= \sqrt{n/\ap}$, its right handed counterpart $\bar w_\al$ must be given as follows:
\beq
\Psi^A \eq \vecb v^\al \\ \bar w_{\dal} \vece \co \bar w_{\dal} \eq \sqrt{\frac{\ap}{n}} \; (D-2) \, \rhob_{\dal}
\label{Dirac1}
\eeq
In the reducible 32 dimensional Dirac spinor space of $SO(1,9)$, the gamma matrices $(\Ga^{\mu})_{A}{}^{B}$ and the charge conjugation matrix ${\cal C}_{AB}$ are both block off-diagonal
\beq
(\Ga^{\mu})_{A}{}^{B} \eq \ccb 0 &\ga^\mu_{\al \dbe} \\ \bar \gamma^{\mu \dal \be}  &0 \cce \co {\cal C}_{AB} \eq \ccb 0 &C_\al {}^{\dbe} \\ C^{\dal}{}_{\be} &0 \cce  
\label{Dirac2}
\eeq
such that their product becomes block diagonal
\beq
(\Ga^{\mu} \, {\cal C})_{AB} \eq \ccb (\ga^\mu \, C)_{\al \be} &0 \\ 0 &(\bar \gamma^{\mu} \, C)^{\dal \dbe} \cce \ .
\label{Dirac2a}
\eeq
The ten dimensional chirality matrix $\Ga_{11}$ is defined by its asymmetric action on the left- and right handed components $v^\al$, $\bar w_{\dbe}$ of a Dirac spinor $\Psi$,
\beq
(\Ga_{11} \, {\cal C})_{AB} \, \Psi^B \eq \vecb - \, C_\al {}^{\dbe} \, \bar w_{\dbe} \\ + \, C^{\dal}{}_{\be} \, v^{\be} \vece \ ,
\label{Dirac3}
\eeq
the charge conjugation matrix ${\cal C}$ is suppressed in the results of subsection \ref{sec:Dirac}. From this, it follows easily how to translate the Weyl spinor contraction of (\ref{f2}) and its special cases into Dirac spinor language $\Psi^A_{2,3} = \left( \begin{smallmatrix} (v_{2,3})^\al \\ (\rhob_{2,3})_{\dal} \end{smallmatrix} \right)$:
\begin{align}
(v_2 \, \ga^\mu \, v_3) \ \ &= \ \ \frac{1}{2} \; \bigl(\Psi_2 \, \Ga^\mu \, (1+\Ga_{11}) \, \Psi_3 \bigr) \notag \\
(v_2 \! \not \! k_3 \, v_3) \ \ &= \ \ \frac{\sqrt{ n_3 }}{2 \sqrt{\ap}} \; \bigl(\Psi_2 \,  (1-\Ga_{11}) \, \Psi_3 \bigr) \label{Dirac4} \\
(v_2 \! \not \! k_2 \, v_3) \ \ &= \ \ - \, \frac{\sqrt{ n_2 }}{2 \sqrt{\ap}} \; \bigl(\Psi_2 \,(1+\Ga_{11}) \, \Psi_3 \bigr) \notag \\
(v_2 \! \not \! k_2\, \gab^\mu \! \not \! k_3  \, v_3) \ \ &= \ \ - \, \frac{\sqrt{n_2 \, n_3 }}{2 \, \ap} \; \bigl(\Psi_2 \, \Ga^\mu \, (1-\Ga_{11}) \, \Psi_3 \bigr) \notag
\end{align}

\end{document}